\documentclass[iop,numberedappendix]{emulateapj}

\usepackage{array} 
\usepackage{hyperref}
\usepackage{amsfonts}
\usepackage{amssymb} 
\usepackage{amsmath}
\usepackage{graphicx}
\usepackage{enumerate}
\usepackage{verbatim}
\usepackage{color}
\usepackage{mathrsfs}
\usepackage{comment} 
\bibpunct{(}{)}{;}{a}{}{,} 

\shorttitle{The radius discrepancy in single and binary stars}
\shortauthors{Spada et al.}

\begin{document}

\title{The radius discrepancy in low mass stars: single vs. binaries}
\author{F.~Spada}
\affil{Leibniz-Institut f\"ur Astrophysik Potsdam (AIP), An der Sternwarte 16, D-14482, Potsdam, Germany}
\email{fspada@aip.de}

\author{P. Demarque}
\affil{Department of Astronomy, Yale University, New Haven, CT 06520-8101, USA}

\author{Y. -C. Kim}
\affil{Yonsei University Observatory and Astronomy Department, Yonsei University, Seoul 120-749, South Korea}

\and

\author{A. Sills}
\affil{Department of Physics and Astronomy, McMaster University, Hamilton, ON, L8S 4M1, Canada }

\begin{abstract}

A long-standing issue in the theory of low mass stars is the discrepancy between predicted and observed radii and effective temperatures.
In spite of the increasing availability of very precise radius determinations from eclipsing binaries and interferometric measurements of radii of single stars, there is no unanimous consensus on the extent (or even the existence) of the discrepancy and on its connection with other stellar properties (e.g. metallicity, magnetic activity).
We investigate the radius discrepancy phenomenon using the best data currently available (accuracy $\lesssim 5\, \%$).
We have constructed a grid of stellar models covering the entire range of low mass stars ($0.1$--$1.25$ $M_\odot$) and various choices of the metallicity and of the mixing length parameter $\alpha$.
We used an improved version of the Yale Rotational stellar Evolution Code (YREC), implementing surface boundary conditions based on the most up-to-date PHOENIX atmosphere models.
Our models are in good agreement with others in the literature and improve and extend the low mass end of the Yale-Yonsei isochrones.
Our calculations include rotation-related quantities, such as moments of inertia and convective turnover time scales, useful in studies of magnetic activity and rotational evolution of solar-like stars.
Consistently with previous works, we find that both binaries and single stars have radii inflated by about $3\, \%$ with respect to the theoretical models; among binaries, the components of short orbital period systems are found to be the most deviant. 
We conclude that both binaries and single stars are comparably affected by the radius discrepancy phenomenon.

\end{abstract}


\def\aj{{AJ}}                   
\def\araa{{ARA\&A}}             
\def\apj{{ApJ}}                 
\def\apjl{{ApJ}}                
\def\apjs{{ApJS}}               
\def\aap{{A\&A}}                
\def\apss{{Ap\&SS}}          
\def\aapr{{A\&A~Rev.}}          
\def\aaps{{A\&AS}}              
\def\mnras{{MNRAS}}             
\def\nat{{Nature}}              
\def\ssr{{Space~Sci.~Rev.}}

\defcitealias{BCAH98}{BCAH98}
\defcitealias{Dotter_ea:2008}{D08}
\defcitealias{Demarque_ea:2004}{Y2}

\section{Introduction}

Low mass stars (i.e., stars of solar mass or lower) are fascinating objects. 
During their main sequence (MS) lifetime, these stars have a convectively unstable region located beneath the surface (and extending all the way to the center for $M \leq 0.3$ $M_\odot$), where large scale magnetic fields are generated through the dynamo mechanism \citep[see, e.g.,][]{Charbonneau:2013}. 
This interior structure is reminiscent of that of the Sun, with which they share many properties: strong surface magnetic fields ($0.01$--$1$ kG, see \citealt{Reiners:2012}); the host of phenomena usually referred to as magnetic activity (e.g. starspots, {\it faculae}, flares, $\dots$), a manifestation of magnetically heated upper atmosphere \citep{Schrijver_Zwaan:2008}; a non-trivial rotational evolution, driven by the braking torque applied to the stellar surface by magnetized winds \citep{Schatzman:1962,Skumanich:1972,Kawaler:1988}.
Furthermore, M type stars ($M\leq 0.6\, M_\odot$) are the most abundant objects in the Galaxy, making up almost $75\%$ in number of its stellar content, and are very promising candidates for the search of exoplanets located in the so-called habitable zone \citep{Kasting_ea:1993,Kopparapu_ea:2013}.
A solid theoretical understanding of the fundamental parameters of low mass stars, leading to precise characterization of these objects from the available observables (e.g., mass--magnitude, mass--radius relations) is therefore very valuable.
 
A long standing issue in the theory of low mass stars is the disagreement between theoretically-derived and observed global parameters.
Discrepancies in radius and effective temperature have been known to exist for a very long time (e.g., \citealt{Hoxie:1973,Lacy:1977}; see also \citealt{Torres_ea:2010} for a review).

The most precise observational constraints on the mass--radius relation of stars have been traditionally provided by the light curve analysis of detached eclipsing binaries (DEBs).
In these systems, masses and radii of the components can be determined from the light curve analysis very accurately (a few percent) and free of additional assumptions.
Until recently, only a handful of such very valuable systems hosting stars in the low mass range were known \citep{Torres_ea:2010}.
However, the situation has improved significantly in the last few years \citep{Feiden_Chaboyer:2012}; moreover, the number of DEBs of known parameters can be expected to further increase in the near future, as a by-product of planet-searching missions, such as {\it Kepler}.

In DEBs, inflated radii and cooler effective temperatures in comparison with the theoretical models are commonly reported (by $5$--$15\%$ and $3$--$5\%$, respectively, see, e.g., \citealt{Torres:2013} and references therein). 
The radius and effective temperature discrepancies roughly compensate each other to give the same luminosity of the model, suggesting a surface origin of the phenomenon. 

Due to unavailability of theoretical models and/or reliable observational constraints on the chemical composition of the stars, fitting of DEBs has often been attempted in the past making use of solar metallicity models only.
Taking into account the metallicity information, \citet{Feiden_Chaboyer:2012} were able to bring the radius discrepancy down to more moderate levels (around or below $4\%$ for their entire sample).

The radius inflation is usually explained as the consequence of an enhanced level of magnetic activity (see, e.g., \citealt{Lopez-Morales:2007,Chabrier_ea:2007}).
Stars in DEB systems can be kept in a regime of fast rotation through the spin-orbit synchronization induced by tidal interactions.
The high rotation rate generates strong magnetic fields via dynamo action \citep{Charbonneau:2013}, which in turn lead to high star spots coverage of the surface and suppression of convection within the star \citep{Gough_Tayler:1966}. 
Both these effects limit the flux of energy throughout the star, which expands in response.  
However, since magnetic activity is enhanced by fast rotation, close-in orbit DEBs, provided they are old enough to have synchronized their rotation periods with the orbital period, should display larger discrepancy than long period systems.
This is not always the case, as both short period, non-discrepant systems (e.g. KOI-126, $P_{\rm orb} = 1.77$ d, \citealt{Carter_ea:2011}) and long period, discrepant ones (e.g. Kepler 16, $P_{\rm orb}=41.1$ d \citealt{Doyle_ea:2011,Winn_ea:2011}) are known. 

The constraints on the fundamental parameters of stars coming from DEBs have recently been complemented by the direct interferometric determinations of stellar radii.
This technique can now reach a precision on the radius comparable to that of light curve analysis in DEBs \citep{Berger_ea:2006,Boyajian_ea:2012a,Boyajian_ea:2012b}; the mass, however, cannot be obtained independently of other measurements (e.g., empirical mass--magnitude relations).
Nevertheless, these measurements can provide very valuable information on whether radius and effective temperature deviations exist in single stars. 
Since no magnetic activity enhancement mechanism via tidal interaction with a companion can be expected in this case, field stars of solar age and approximately solar mass should show very moderate or no radius discrepancy.
A $10\%$ radius deviation with respect to the models of \citet{BCAH98} has been found by \citet{Berger_ea:2006}; however, this result has been challenged by \citet{Demory_ea:2009}.
  
We present a grid of models in the mass range $0.1$--$1.25\; M_\odot$ calculated for various choices of the metallicity and the mixing length parameter.
One of our goals is to extend and update the low mass end of the Yale-Yonsei (Y$^2$) isochrones \citep{Yi_ea:2001,Demarque_ea:2004}.
Our present calculations rely on up-to-date input physics, capable of reproducing accurately the conditions of stars at the low mass end of the range (e.g., high density/low temperature EOS, surface boundary conditions based on realistic model atmospheres).

This work is also intended as the starting point of a series of papers on related topics.
One planned application is the study of the rotational evolution of solar-like and very low mass stars and in particular the physical mechanisms providing rotational coupling in stellar interiors \citep[see, e.g.,][]{Spada_ea:2010,Spada_ea:2011}. 
We have thus calculated rotation-related quantities, such as the moments of inertia of the radiative and convective zones of the stars, as well as the convective turnover time scale. 
The latter is a global measure of the efficiency of convection and has been often used as a means to connect stellar rotation and activity (\citealt{Noyes_ea:1984}; see also \citealt{Barnes_Kim:2010,Barnes:2010}).

Secondly, we will investigate the effect of variable convective efficiency in stellar 1D models.
In conventional stellar models, convection is described by means of the classical mixing length theory (MLT, \citealt{BV58}). 
The so-called mixing length parameter $\alpha$, the ratio between the mixing length itself and the local pressure scale height, is a measure of the efficiency of convection, and, being a free parameter in the MLT, is typically calibrated on the Sun. 
There is increasing evidence, however, that the use of a solar-calibrated mixing length is not warranted for other stars, and that $\alpha$ should be tied to stellar properties instead \citep{Bonaca_ea:2012}. 
Theoretical investigations of these effects, with the goal of formulating prescriptions that could be incorporated in 1D models, are well under way \citep[e.g.,][]{Tanner_ea:2013}. 

A reduction of convective efficiency has also been proposed as the main global effect on stellar structure of the presence of magnetic fields.
This effect has been taken phenomenologically into account by, e.g., \citet{Chabrier_ea:2007}, using an effective value of $\alpha$, reduced in comparison with its solar-calibrated value.
A more quantitative approach, where the reduction of $\alpha$ is connected with the vertical magnetic field intensity according to the formulation of \citet{Gough_Tayler:1966}, has been adopted by \citet{MacDonald_Mullan:2012,MacDonald_Mullan:2013}.
Finally, the formalism developed by \citet{Lydon_Sofia:1995} in an early attempt at a self-consistent $1D$ modeling of the Sun with magnetic fields was recently applied to the DEB system EF Aquarii by \citet{Feiden_Chaboyer:2012b}.  
Further theoretical efforts along these lines are a natural continuation of our present investigation on the radius discrepancy problem.
%

Our grid will also be useful in those studies where knowledge of a precise mass--radius theoretical relationship is crucial. 
For example, in the characterization of exoplanet-hosting systems, the accuracy in the determination of the mass and the radius of the planets is contingent on that of the hosting star \citep[e.g.][]{Henry:2004}.

In the present paper, we investigate the (theoretical vs. observational) relationships among the fundamental parameters of low mass stars. 
Our aim is to shed light on the nature of the physical mechanisms producing the radius and effective temperature discrepancies in low mass stars, by disentangling the intrinsic disagreement from that caused by other effects (e.g., age and/or metallicity uncertainties), and by ascertaining whether the observational data on binaries and single stars (from DEBs and interferometry, respectively) can be reconciled within a consistent paradigm.
To this end, we compare our models with the two largest samples currently available that contain low mass stars and satisfy a high accuracy criterion (i.e. the compilation of DEBs by \citealt{Feiden_Chaboyer:2012} and the interferometric measurements by \citealt{Boyajian_ea:2012b}, respectively).

\section{The models}

\subsection{Description of the code}

All the models were calculated using the Yale Rotational stellar Evolution Code (YREC) in its non-rotational configuration.
Compared with the standard version of YREC \citep[see, e.g.,][]{Demarque_ea:2008}, the code used in this work contains some major improvements.
The most important ones concern the treatment of the equation of state and the use of outer boundary conditions based on updated non-grey atmospheric models (see Sect.~\ref{sec:surfbc} below). 
These improvements are particularly significant when modelling low mass objects, i.e. for $M \leq 0.6 \, M_\odot$. 
Our present calculations, therefore, supersede the Y$^2$ models in this range of mass.
 
\subsubsection{Standard input physics in YREC}
The details of the input physics are as follows. 
We used \citet{Ferguson_ea:2005} opacities at low temperatures, and the OPAL Rosseland opacities at high temperatures \citep{Iglesias_Rogers:1996}; the Equation of State (EOS) is the OPAL~2005 EOS \citep{Rogers_Nayfonov:2002}. 
The energy generation rates are calculated according to the prescription of \citet{Bahcall_Pinsonneault:1992}; diffusion of helium and heavy elements is taken into account,  with the diffusion coefficients calculated according to the prescription of \citet{Thoul_ea:1994}. 
Convection is described with the mixing length theory \citep[MLT;][]{BV58}; the value used for the MLT parameter $\alpha$ (the mixing length scaled over the pressure scale height) is discussed below (see Sect.~\ref{thegrid}).
In all calculations, we adopt the \citet{Grevesse_Sauval:1998} value of the solar metallicity, $(Z/X)_\odot = 0.0230$ and an $\alpha$-enhancement $[{\rm \alpha/Fe}]=0$.
Convective core overshooting is not necessary in modeling stars in the mass range consider here and therefore it is not taken into account.

\subsubsection{Equation of State and surface boundary conditions for low mass models}
\label{sec:surfbc}
The standard treatment of the atmospheric boundary conditions in YREC relies on the specification of a temperature-optical depth ($T$--$\tau$) relationship \citep[typically, the one of the classical Eddington grey atmosphere model or the empirical one from][]{KrishnaSwamy:1966}.
However, although this approximation is adequate for stars of solar mass or higher, it can lead to significant errors in the determination of the stellar radius and effective temperature for stars of mass $\lesssim 0.3~M_\odot$ (\citealt{Chabrier_Baraffe:1997}; see also \citealt{Spada_Demarque:2012}).
Since the mass range we wish to investigate extends below this limit, we incorporated in the standard version of YREC, described so far, the improvements discussed in \citet{Sills_ea:2000}:
the use of the SCVH EOS \citep{Saumon_ea:1995} in the low temperature--high density regime, and of the surface boundary conditions derived from PHOENIX model atmospheres \citep{Hauschildt_ea:1999}.

In the current YREC implementation, the OPAL and SCVH EOS are ramped at $T = 5000$ K. 
For the surface boundary conditions, the value of the gas pressure at the stellar photosphere is interpolated from a table compiled from the PHOENIX atmosphere models.
To achieve overall consistency within the grid, we used the surface boundary conditions derived from PHOENIX throughout the entire mass range covered.
The original input tables used by \citet{Sills_ea:2000}, generated from NextGen models \citep{Hauschildt_ea:1999}, were updated to include the new BT-Settl models \citep{Allard_ea:2011}, available from F. Allard's web page \footnote{\texttt{http://perso.ens-lyon.fr/france.allard/}}, and to account for non-solar metallicities.

As the BT-Settl models are only available for either the \citet{Grevesse_Noels:1993} or the \citet{Asplund_ea:2009} chemical compositions, complete consistency cannot be achieved with the composition we adopted for the interior calculations (i.e. that of \citealt{Grevesse_Sauval:1998}). 
We have used the BT-Settl atmosphere models with \citet{Grevesse_Noels:1993} mixture, which is the closest to our choice for the interior. 
The interior and atmosphere models are matched according to the value of ${\rm [Fe/H]}$ (i.e., they will not have exactly the same value of $Z/X$ or of $Z$). 
The relative difference between the \citet{Grevesse_Noels:1993} and the \citet{Grevesse_Sauval:1998} value of $(Z/X)_\odot$ is however, quite small (about $6\%$). 
Moreover, the difference in $Z$ will only have second order effects on the stellar models, as a change in the atmospheric $Z$ will result in a change in the value of the photospheric pressure used in the outer boundary conditions, and this, in turn, will affect the stellar model.

\begin{figure}
\begin{center}
\includegraphics[width=0.45\textwidth]{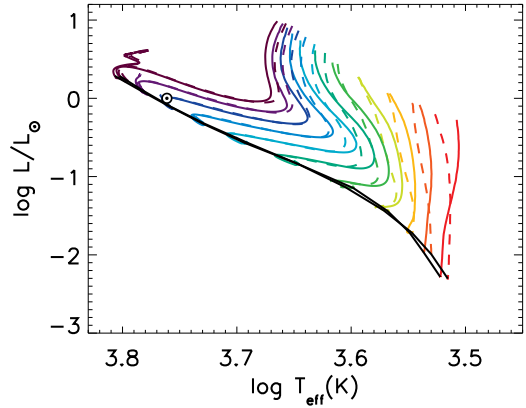}
\caption{Impact of different atmosphere models: NextGen \citep{Hauschildt_ea:1999} vs. BT-Settl \citep{Allard_ea:2011} models. 
Solid lines: evolutionary tracks (in color) and $0.5$ Gyr isochrone (black) for $0.2 \leq M/M_\odot \leq 1.2$, $[{\rm Fe/H}]=0$, solar-calibrated $\alpha$($=1.875$), with surface boundary conditions taken from the BT-Settl atmospheric models. Dashed lines: the same, but using NextGen atmospheric models.}
\label{fig:NGvsBT}
\end{center}
\end{figure}

A comparison of the evolutionary tracks in the HR diagram obtained using the surface boundary conditions from NextGen and from BT-Settl models is shown in Figure~\ref{fig:NGvsBT}.
Although the effects on the isochrone shown are rather modest, the differences in the pre-main sequence (PMS) tracks are non-negligible, especially for the lowest masses.  

\subsubsection{Standard solar model calibration}
\label{sec:ssm}
Using the input physics described so far, we calculated a standard solar model, i.e. we calibrated the initial helium content $Y_0$ and the MLT parameter $\alpha$ in order for a $1~M_\odot$ model to match the solar radius and the solar luminosity at the age of $4.57$ Gyr, having set the photospheric value of $(Z/X)_\odot=0.0230$ \citep{Grevesse_Sauval:1998}. 
Our best solar model has the initial composition $Y_{0,\odot}=0.27783$, $Z_{0,\odot}=0.018811$, and $\alpha_\odot = 1.8750$; other details are listed in Table~\ref{tab:ssm}.

\begin{table}
\begin{center}
\caption{Standard solar model.}
\begin{tabular}{ccc}
\hline
\hline
Parameter & Adopted$^\dag$ & Model \\
\hline 
Age (Gyr) & $4.57$ & - \\
Mass (g) & $1.9891 \cdot 10^{33}$ & - \\  
$R$ (cm) & $6.9598 \cdot 10^{10}$ & $\log({R/R_\odot}) = 5 \cdot 10^{-8}$ \\
$L$ (erg/s) & $3.8418 \cdot 10^{33}$ & $\log( {L/L_\odot}) =4 \cdot 10^{-8}$ \\ 
${R_{BCZ}/R_\odot}$ & $0.713$  &  $0.715$ \\
$({Z/X})_{\rm surf}$ & $0.0230$ & $0.0230$ \\
\hline
\end{tabular}
\label{tab:ssm}
\end{center}
$^\dag$ See, e.g., \cite{Basu_Antia:2008}.
\end{table}%

\subsection{Parameters of the grid}
\label{thegrid}

Our grid of models covers the mass range $0.10~M_\odot \leq M \leq 1.25~M_\odot$ with increments of $0.05~M_\odot$.
This suffices to encompass the entire range of solar-like stars (i.e., stars with a subsurface convection zone and an inner radiative region), as well as stellar objects below the transition to the fully convective regime.
Modelling the brown dwarf regime is beyond the scope of the present work.

The initial chemical composition of the models is assigned as follows. 
We have calculated models for five different values of the metallicity (scaled to the solar vaue): $[{\rm Fe/H}] = +0.3$, $0.0$, $-0.5$, $-1.0$, $-1.5$, where $\left(\frac{Z}{X}\right)_* = \left(\frac{Z}{X}\right)_\odot \, 10^{[{\rm Fe/H}]}$. 
The helium mass fraction $Y$ is scaled with the heavy elements mass fraction $Z$ according to the relation $Y = Y_p + (\frac{\Delta Y}{\Delta Z})\, Z$, where $Y_p = 0.25$ is the primordial helium abundance \citep{Cyburt_ea:2008} \footnote{See also the review and references at \url{http://www.pas.rochester.edu/~emamajek/memo_Yp.html}.}.
In all the calculations, the enrichment parameter $(\frac{\Delta Y}{\Delta Z}) $ is equal to $1.48$, based on our standard solar model calibration (see Section~\ref{sec:ssm}). 

 The initial composition of the models, given by:
 \begin{eqnarray*}
X = \frac{1 - Y_p}{1 +   \left(\frac{Z}{X} \right)_*\left[1 +  \frac{\Delta Y}{\Delta Z}  \right]} \ ; \ \ \ \ 
Z = \left(\frac{Z}{X}\right)_* \, X,
\end{eqnarray*}
is listed in Table~\ref{tab:comp}.
\begin{table}[htdp]
\caption{Initial compositions of the models in the grid.}
\begin{center}
\begin{tabular}{cccc}
\hline
\hline
$[{\rm Fe/H}]$ & $Y = 0.25 + 1.48\, Z$ & $X$ & $Z$ \\
\hline
$+0.3$ & $0.29573$ & $0.67336$  & $0.03090$ \\
$\,0.0$  & $0.27415$ & $0.70952$ & $0.01631$ \\
$-0.5$ &  $0.25793$ & $0.73671$ & $0.00535$ \\
$-1.0$ &  $0.25253$ & $0.74574$ & $0.00171$ \\
$-1.5$ &  $0.25080$ & $0.74865$ & $0.00054$ \\
\hline 
\end{tabular}
\end{center}
\label{tab:comp}
\end{table}%

The solar-calibrated value of the MLT parameter $\alpha$ is equal to $\alpha_\odot=1.875$ (see Sect.~\ref{sec:ssm}); however, in order to investigate its impact on the structure and on the evolutionary tracks, we have also calculated models with $\alpha=0.50$, $1.00$, and $3.00$.
In general, reducing the value of $\alpha$ results in models with larger radii and cooler effective temperature; the sensitivity to $\alpha$ decreases approximately exponentially with decreasing mass.
An effective value of the MLT parameter $\alpha$ has been used in the past by many authors to mimic the reduced convection efficiency induced by magnetic fields \citep[e.g.][]{Chabrier_ea:2007}. 
A detailed analysis of this effect in the context of the radius discrepancy problem, discussed in Sec.~\ref{sec:mrl}, is deferred to future work.

The initial models have homogeneous composition and a polytropic structure. 
They are based on the birth line models constructed by Dr. Sydney Barnes at an early stage of the Y$^2$ project \citep[see][]{Yi_ea:2001}. 
Each track is evolved through the PMS and MS, until either the bottom of the red giant branch or the age of $13$ Gyr is reached.

\begin{figure}[htdp]
\begin{center}
\includegraphics[width=0.45\textwidth]{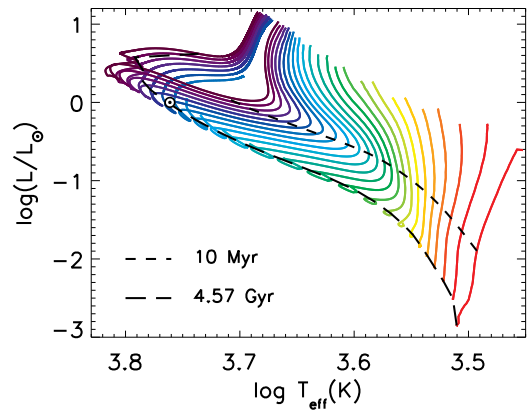}
\caption{Theoretical tracks in the HRD for models of mass $0.1 \leq M/M_\odot \leq 1.25$ (in steps of $0.05$ $M_\odot$), along with $10$ Myr and $4.57$ Gyr isochrones, for $[{\rm Fe/H}] = 0.0$, $\alpha = 1.875$. The position of the present-day Sun is also shown.}
\label{fig:hrsolar}
\end{center}
\end{figure}

The evolutionary tracks in the Hertzsprung-Russel diagram (HRD) for a representative subset of our grid ($[{\rm Fe/H}]=0.0$, $\alpha = \alpha_\odot$) are shown in Figure \ref{fig:hrsolar}.

\subsection{Description of the tracks and isochrones}
All the models in our grid are available for download from the Y$^2$ project web page: \texttt{http://www.astro.yale.edu/demarque/yyiso.html}.
We provide evolutionary tracks and isochrones for the models in the range $0.1 \leq M/M_\odot \leq 1.25$, with $\alpha = 0.5, \, 1.0, \, 1.875, \, 3.0$, and chemical composition as in Table~\ref{tab:comp}.
The evolutionary tracks cover the PMS and the post-MS up to $13$ Gyr and/or the sub giant phase (if present). 
The basic quantities are given in the \texttt{.track1} files. 
The rotation-related quantities, discussed in Appendix~\ref{app:rotevol}, are given in the \texttt{.track2} files.

The isochrones also contain synthetic magnitudes, obtained from the $T_{\rm eff}$--color transformations of \citet{Lejeune_ea:1998} and \citet{VandenBerg_Clem:2003} (\texttt{.iso1} and \texttt{.iso2} files, respectively). 
For a discussion of the  $T_{\rm eff}$--color transformations, see Appendix~\ref{app:colteff}).  

The content of the \texttt{.track1}, \texttt{.track2}, \texttt{.iso1}, and \texttt{.iso2} files is illustrated in Tables~\ref{tab:tr1}--\ref{tab:iso}.

\begin{table*}[htdp]
\caption{Content of the \texttt{.track1} files.}
\begin{center}
\begin{tabular}{ccccccccccccc}
\hline
\hline
Age & $X_c$ & $Y_c$ & $Z_c$ & $\log L/L_\odot$ & $\log R/R_\odot$ & $\log g$ & $\log T_{\rm eff}$ & $U_{\rm grav}$ & $X_{\rm env}$ & $Z_{\rm env}$ & $\log T_c$ & $\eta$ \\
(1) & (2) & (3) &(4) &(5) &(6) &(7) &(8) &(9) &(10) &(11) &(12) &(13) \\
\hline
\dots & \dots & \dots & \dots & \dots & \dots & \dots & \dots & \dots & \dots & \dots & \dots & \dots \\
\end{tabular}
\end{center}
(1): Age [Gyr]; (2): hydrogen mass fraction in the center; (3): helium mass fraction in the center; (4): mass fraction of heavy elements in the center; (5): luminosity $[L_\odot]$; (6): radius $[R_\odot]$; (7) surface gravity [g/cm$^2$]; (8): effective temperature [K]; (9): fraction of energy produced by gravitational contraction [$\%$]; (10) hydrogen mass fraction at the surface; (11): mass fraction of heavy elements at the surface; (12): central temperature [K]; (13) degeneracy parameter. 
\label{tab:tr1}
\end{table*}%
\begin{table*}[htdp]
\caption{Content of the \texttt{.track2} files.}
\begin{center}
\begin{tabular}{cccccccc}
\hline
\hline
Age & $M_{\rm core}$ & $M_{\rm envp}$ & $R_{\rm envp}/R_*$ & $\tau_c$ & $I_{\rm core}$ & $I_{\rm envp}$ & $I_{\rm tot}$ \\
(1) & (2) & (3) &(4) &(5) &(6) &(7) &(8) \\
\hline
\dots & \dots & \dots & \dots & \dots & \dots & \dots & \dots \\
\end{tabular}
\end{center}
(1): Age [Gyr]; (2): mass of the convective core (if present) $[M_*]$; (3): mass of the convective envelope $[M_*]$; (4): fractional radius at the bottom of the external convective envelope $[R_*]$; (5): global convective turnover time scale [days]; (6): moment of inertia of the convective core (if present) [g cm$^2$]; (7) moment of inertia of the external convective envelope [g cm$^2$]; (8): total moment of inertia of the star [g cm$^2$]. 
\label{tab:tr2}
\end{table*}%
\begin{table*}[htdp]
\caption{Content of the \texttt{.iso1} and \texttt{.iso2} files.}
\begin{center}
\begin{tabular}{ccccccccc}
\hline
\hline
Mass  & $\log T_{\rm eff}$ & $\log L/L_\odot$ & $\log g$ & $M_V$ & $U-B$ & $B-V$ & $V-R$ & $R-I$ \\
(1) & (2) & (3) &(4) &(5) &(6) &(7) &(8) & (9) \\
\hline
\dots & \dots & \dots & \dots & \dots & \dots & \dots & \dots \\
\end{tabular}
\end{center}
(1): Mass $[M_\odot]$; (2): effective temperature [K]; (3): luminosity $[L_\odot]$; (4): surface gravity [g cm$^2$]; (5): absolute $V$ magnitude [mag]; (6): $U-B$ color index; (7) $B-V$ color index; (8): $V-R$ color index; (9): $R-I$ color index . 
\label{tab:iso}
\end{table*}%

\begin{figure*}[htdp]
\begin{center}
\includegraphics[width=0.9\textwidth]{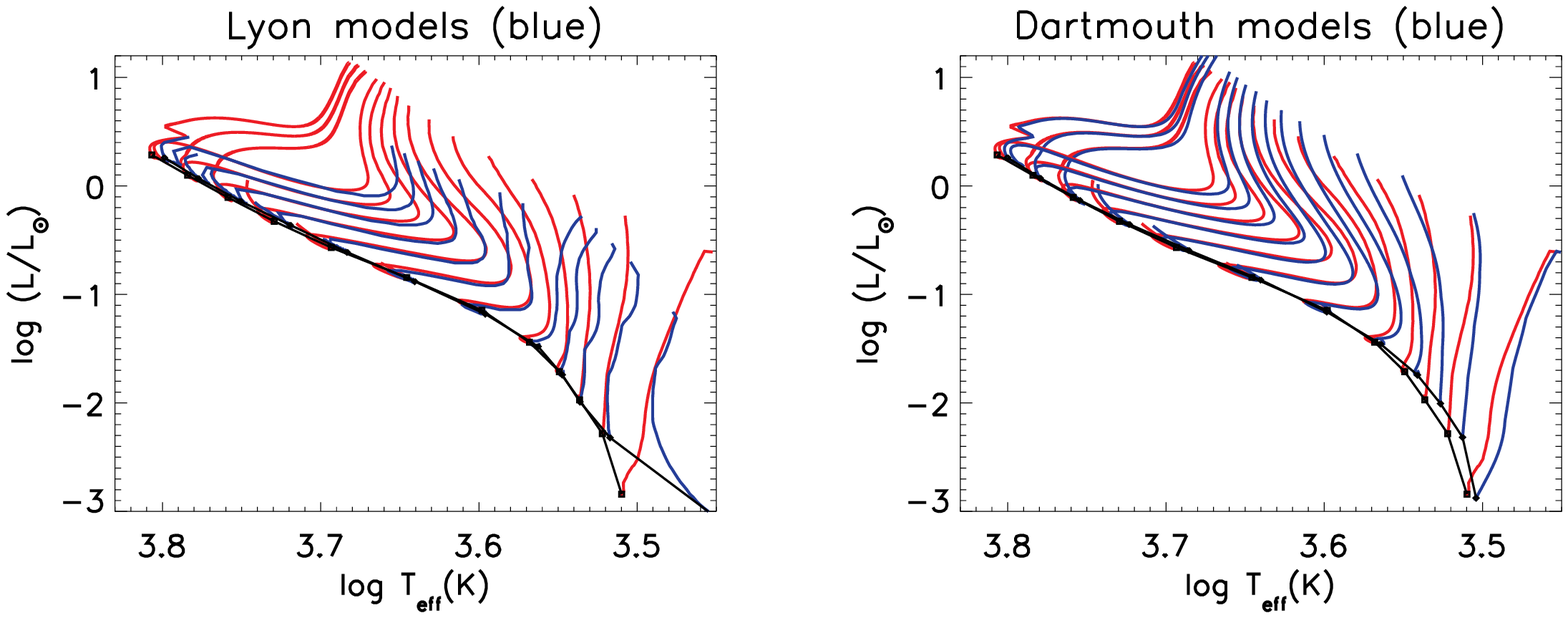}
\includegraphics[width=0.9\textwidth]{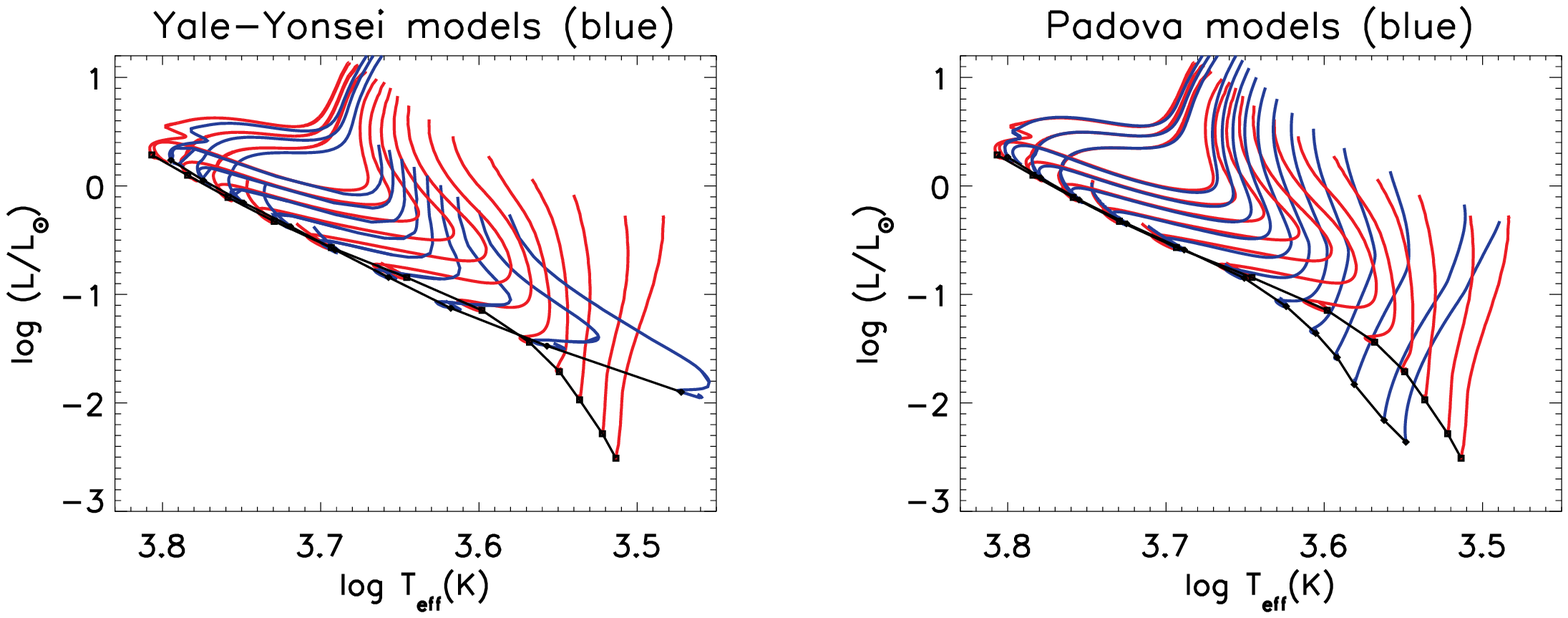}
\caption{Comparison between our models (solar metallicity, solar-calibrated $\alpha$ and $M = 0.1$, $0.2$, $\dots 1.2$ $M_\odot$; in red) and other sets of publicly available tracks and isochrones (blue). A $0.5$ Gyr isochrone is shown in each panel.
Upper left: Lyon models (\citealt{BCAH98}; note that the tracks are not evolved all the way through the subgiant branch); upper right: Dartmouth \citep{Dotter_ea:2008}; lower left: Yale-Yonsei (\citealt{Demarque_ea:2004}; note that the lowest mass available is $0.4\, M_\odot$); lower right: Padova \citep{Bressan_ea:2012}.}
\label{fig:compmod}
\end{center}
\end{figure*}

\subsection{Comparison with other models}
\label{sec:b98d08}

In this section, we compare our models with those by the Lyon, Dartmouth, Yale-Yonsei, and Padova groups \citep[][respectively]{BCAH98,Dotter_ea:2008,Demarque_ea:2004,Bressan_ea:2012}. 
The comparison between tracks (in our mass range, $0.1 \lesssim M/M_\odot \lesssim 1.25$) and isochrones, all calculated for solar metallicity and solar-scaled MLT parameter, is shown in Figure~\ref{fig:compmod}.

\begin{description}
\item[Lyon models ($\alpha = 1.900$, $Z=0.0172$)] The models by \citet{BCAH98} cover the PMS phase, but do not go all the way through the subgiant branch. 
The largest difference with our models is found for the $0.1 \; M_\odot$ track; however, this mass is close to the lower limit of validity of our current configuration of YREC.
These authors used the SCVH EOS and the NextGen version of the PHOENIX atmosphere models in their surface boundary conditions. 
The SCVH EOS is a pure H and He EOS, specifically designed to take into account non-ideal effects due to high density in low mass stars; it has been shown by \citet{Chabrier_Baraffe:1997} to be in very good agreement with the MHD EOS \citep{Hummer_Mihalas:1988,Mihalas_ea:1988,Dappen_ea:1988} in the common range of validity (for a comparison of MHD and OPAL EOS, see \citealt{Trampedach_ea:2006}). 
Note that our version of YREC uses the SCVH EOS only in the low temperature/high density regime.
The effects of the different versions of the PHOENIX atmospheric models (i.e. NextGen vs. BT-Settl), on the other hand, have been discussed in Sec.~\ref{sec:surfbc}. 
\item[Dartmouth models  ($\alpha = 1.938$, $Z=0.0188$)] The Dartmouth Stellar Evolution Program (DSEP) is a distant relative of YREC. 
The two most important differences between the Dartmouth models and our calculations are in the choice of the EOS and the atmospheric boundary conditions. 
The latest version of DSEP \citep[see][and references therein]{Dotter_ea:2007,Dotter_ea:2008} uses NextGen PHOENIX atmospheres (see Sect.~\ref{sec:surfbc}), and the FreeEOS\footnote{For more information on the FreeEOS, consult the web page: \texttt{http://freeeos.sourceforge.net/}. }.  
This set of models is probably the one in closest agreement with our calculations.
\item[Y$^2$ models  ($\alpha = 1.743$, $Z=0.0200$)]
One of the aims of this work was to revise the Y$^2$ isochrones at low masses.
The Y$^2$ models \citep{Yi_ea:2001,Kim_ea:2002,Demarque_ea:2004} cover the mass range from $0.4$ to $5.0\; M_\odot$; convective core overshooting is taken into account.  
The models were constructed with a pre-\citet{Grevesse_Sauval:1998} solar mixture (i.e. with $(Z/X)_\odot = 0.0245$), and the OPAL 2001 EOS. 
Moreover, only helium diffusion is present in Y$^2$ models. 
However, the most significant difference with our current calculations is in the treatment of the atmosphere, which was based on a grey $T$--$\tau$ relation, unsuitable for $M\leq 0.6\, M_\odot$ (their lowest mass track has $M\leq 0.4\, M_\odot$).
The large differences in Figure~\ref{fig:compmod} for the tracks with lowest masses are therefore not unexpected, while a much better agreement can be observed moving towards the high mass end of the range.
\item[Padova models  ($\alpha = 1.740$, $Z=0.0190$)]
In this comparison we use the latest version of the PARSEC models \citep{Bressan_ea:2012}.
Their calculations make use of the FreeEOS; surface boundary conditions are based on the Eddington grey $T-\tau$ relation; convective core overshooting is taken into account.
Both the tracks and the isochrone shown here are in good agreement with ours for masses larger than $0.6\, M_\odot$.
At lower masses, the use of grey atmospheres results in effective temperatures systematically larger than ours, as expected (see discussion in Sect.~\ref{sec:surfbc}).
\end{description}

In conclusion, the comparisons shown here demonstrate the sensitivity of the low mass models to the surface boundary conditions and, to a lesser extent, to the EOS.
This sensitivity lessens moving towards higher masses.
Our isochrones and tracks are thus in good agreement with those of \citet{BCAH98} and \citet{Dotter_ea:2008}, who adopted similar choices for these pieces of input physics.

\begin{figure}
\begin{center}
\includegraphics[width=0.45\textwidth]{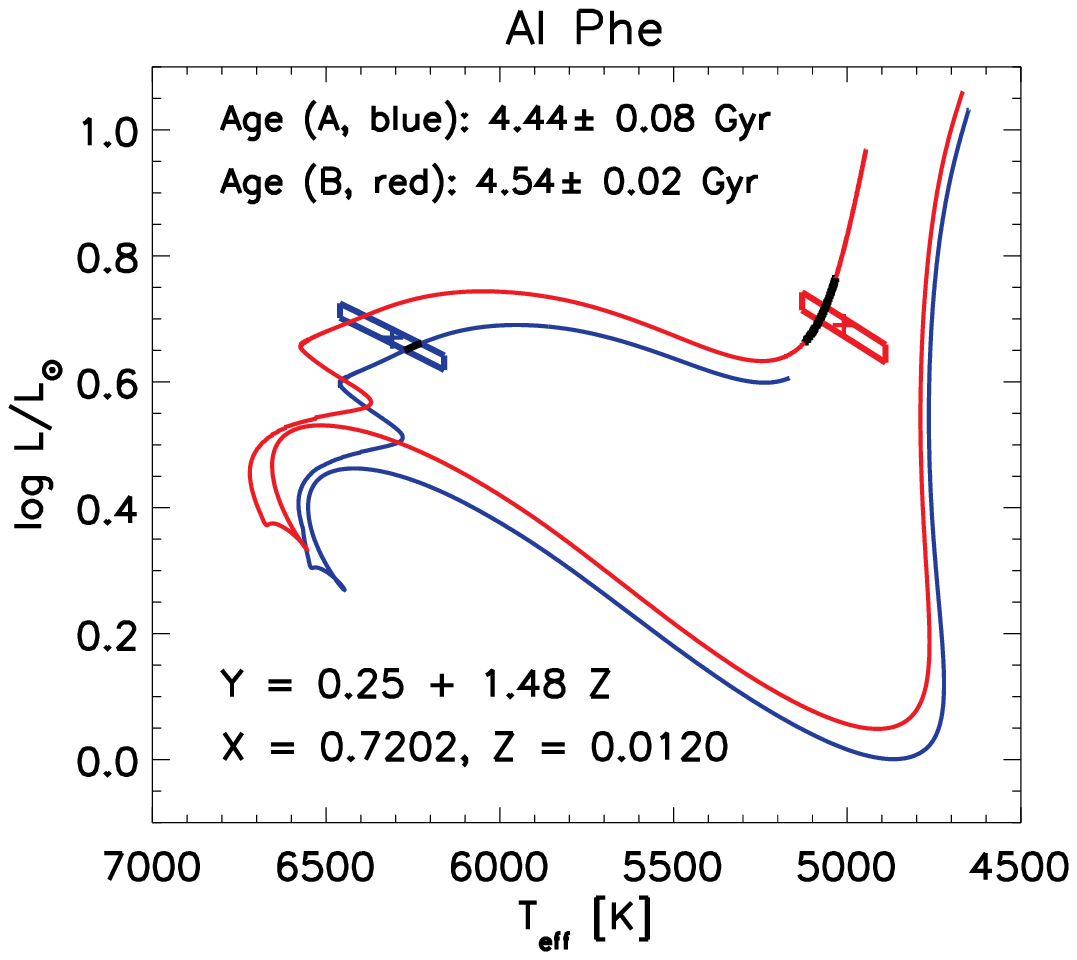}
\caption{Observed properties of AI Phe along with theoretical tracks. Observed positions in the HRD (plus signs), $1\sigma$ uncertainties (boxes), and theoretical tracks are shown in red for the cooler (primary, $B$) component and in blue for the hotter (secondary, $A$) component.
 The portion of each track compatible with the best age within $\pm 1\%$ is shown as a heavier black line, to illustrate the different evolutionary time scales.}
\label{fig:aiPhe}
\end{center}
\end{figure}

\subsection{AI Phoenicis: a stellar evolution benchmark}

Binary systems with accurate determination of the metallicity and of the physical parameters of both components (masses, radii, effective temperatures) provide very valuable benchmarks for stellar evolution models.
One of the best examples is the evolved system AI Phe (\citealt{Andersen_ea:1988}; see also \citealt{Torres_ea:2010}). 

The metallicity of the system has been determined as slightly below solar ($[{\rm Fe/H}]=-0.14\pm 0.1$). 
The masses of the two components, known with a relative accuracy of $0.4\%$ (see Table~\ref{tab:aiPhe}), are sufficiently separate to give rise to a significant difference in the evolutionary time scales.
As a consequence, a non-trivial test for stellar models is whether the best-fitting models give consistently the same age for both components.
\begin{table}
\caption{Measured physical parameters of the system AI Phe, from \citet{Andersen_ea:1988}.}
\begin{center}
\begin{tabular}{ccc}
 \hline
 \hline
 & A (hotter) & B (cooler) \\
 \hline
 $M/M_\odot$ & $1.1954 \pm 0.0041$ & $1.2357 \pm 0.0045$ \\
 $R/R_\odot$ & $1.816 \pm 0.024$ & $2.930 \pm 0.048$ \\
$T_{\rm eff} \; [K]$ & $6310 \pm 150 $ & $5010 \pm 120$ \\
$\log  L/L_\odot$ & $0.67 \pm 0.04$ & $0.69 \pm 0.04$ \\
$[{\rm Fe/H}]$ & \multicolumn{2}{c}{$-0.14 \pm 0.1$} \\
distance [pc]& \multicolumn{2}{c}{$173\pm 11$} \\
 \hline 
\end{tabular}
\end{center}
\label{tab:aiPhe}
\end{table}%

Figure~\ref{fig:aiPhe} shows how our models perform in this test.
The tracks shown in the Figure were calculated using the observed metallicity and masses (listed in Table~\ref{tab:aiPhe}), our solar-calibrated $\alpha$ parameter, and the initial helium content fixed by our enrichment parameter $\Delta Y/ \Delta Z = 1.48$.
As in all the models discussed in the rest of the paper, core overshooting is not taken into account.
%

We estimate the age of each component as the age of the model whose radius coincides with its observed value.
We obtain ages of $4.44$ Gyr and $4.54$ Gyr, respectively, for the $A$ and $B$ component, compatible with each other at the $2\%$ level.
This gives an age of the system in between those obtained by \citet{Torres_ea:2010} using the Y$^2$ isochrones \citep{Demarque_ea:2004} and the Victoria models \citep{VandenBerg_ea:2006}.

\begin{figure}
\begin{center}
\includegraphics[width=0.45\textwidth]{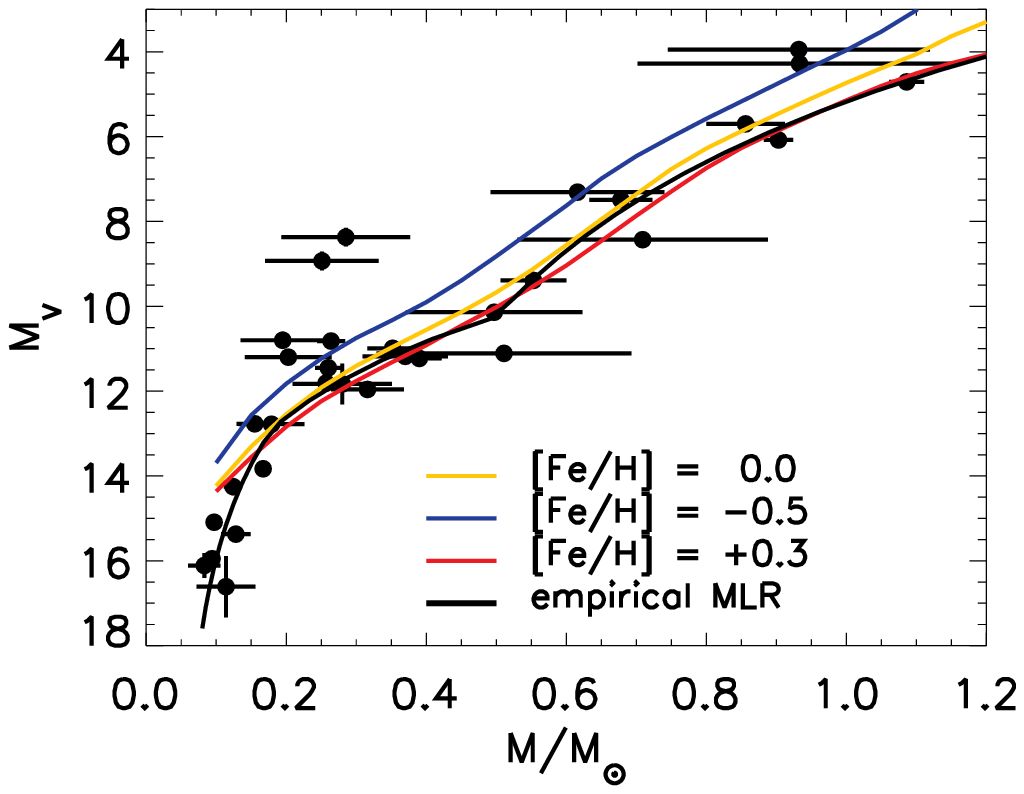}
\includegraphics[width=0.45\textwidth]{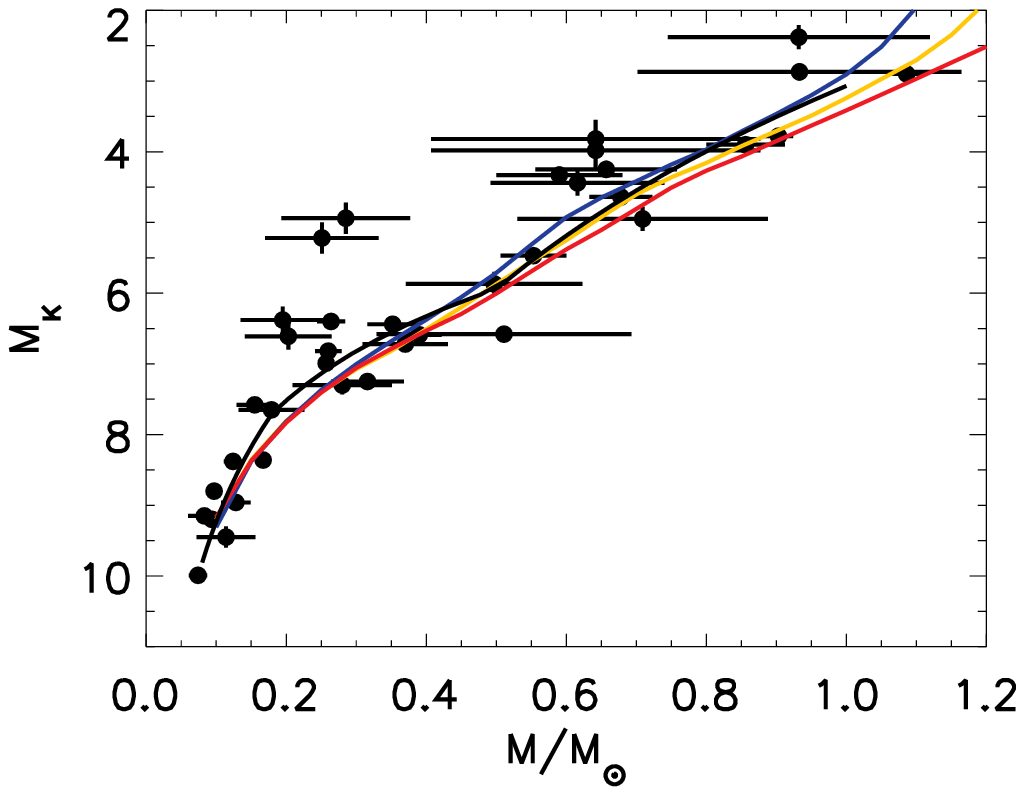}
\caption{Synthetic mass-magnitude relations (upper panel: $V$ magnitude, lower panel: $K$ magnitude) at $5$ Gyr, for models with solar-calibrated $\alpha$ and $[{\rm Fe/H}] = 0.0,\, -0.5$, and $+0.3$ (yellow, blue and red lines, respectively). The data and the empirical MLR (black lines) are from \citet{Henry_McCarthy:1993}.}
\label{fig:hmc93}
\end{center}
\end{figure}

\subsection{Empirical mass--magnitude relations}
\label{sec:emp_masslum}

Empirical mass--luminosity relations (MLR) in infrared and optical wavelengths have been established by \cite{Henry_McCarthy:1993} using data from speckle interferometry of binaries and from the analysis of eclipsing binary systems \citep[see also][]{Henry:2004, Henry_ea:2006}.
The MLRs were obtained through a fitting procedure which does not take into account metallicity and age differences.
The data and the empirical MLRs are compared with the synthetic mass-magnitude relations derived from our models in Figure~\ref{fig:hmc93}.
The Figure shows $5$ Gyr isochrones with $[{\rm Fe/H}]=0.0$, $[{\rm Fe/H}]=-0.5$, $[{\rm Fe/H}]=+0.3$ and $\alpha = 1.875$; the $M_V$ and $M_K$ absolute magnitude were obtained using the \citet{Lejeune_ea:1998} conversions.  
Notably, the MLR in the $K$ band is much less sensitive to metallicity compared to the one in the optical band \citep[see also][]{BCAH98}.

The $K$-band MLR is particularly important in the following, since it has been used by \citet{Boyajian_ea:2012b} to estimate the mass for the stars in their sample.
They assigned a fixed $10\%$ error to the masses determined in this way; this is the largest source of uncertainty in the interferometric sample.

\section{Fundamental parameters of low mass stars}
\label{sec:mrl}

In the following, we discuss in detail the theoretical relations among global stellar properties derived from our models and their agreement with the observational data.

\subsection{The data}

To provide meaningful constraints on the models, an accuracy of $5\%$ or lower on the measured stellar parameters is required \citep{Torres_ea:2010}. 
Currently, the available observations which satisfy this requirement are from two main sources: light curve analysis of DEBs and interferometric measurement of stellar radii.

DEBs provide a very tight constraint on the mass--radius relationship, since both quantities can be determined directly, usually with an accuracy within a few percent; knowledge of other parameters of the system, such as age or metallicity, or of the components, such as effective temperatures or luminosities, although very valuable, is rarely available or equally accurate.
The DEB sample studied by \citet{Feiden_Chaboyer:2012} is the largest available to date satisfying the $5\%$ quality criterion and containing systems with at least one component of mass below $ 0.7\,M_\odot$.
For a few systems in this sample, some information constraining the age and the chemical composition is also available.  

On the other hand, very accurate interferometric measurements of stellar parameters for stars in the solar mass range have been recently reported by \citet{Boyajian_ea:2012b}.
Using direct angular diameter measurements and bolometric fluxes derived from SED fitting in conjunction with {\it Hipparcos} parallaxes, these authors have derived luminosities, radii, and effective temperatures for a sample of single stars in the solar neighborhood, with an accuracy that satisfies the $\leq 5\%$ criterion. 
They also provide literature values of the metallicity and mass estimates from the \citet{Henry_McCarthy:1993} MRL in the $K$ band (an error of $10\%$ is quoted for the masses derived in this way).

\begin{figure}
\begin{center}
\includegraphics[width=0.45\textwidth]{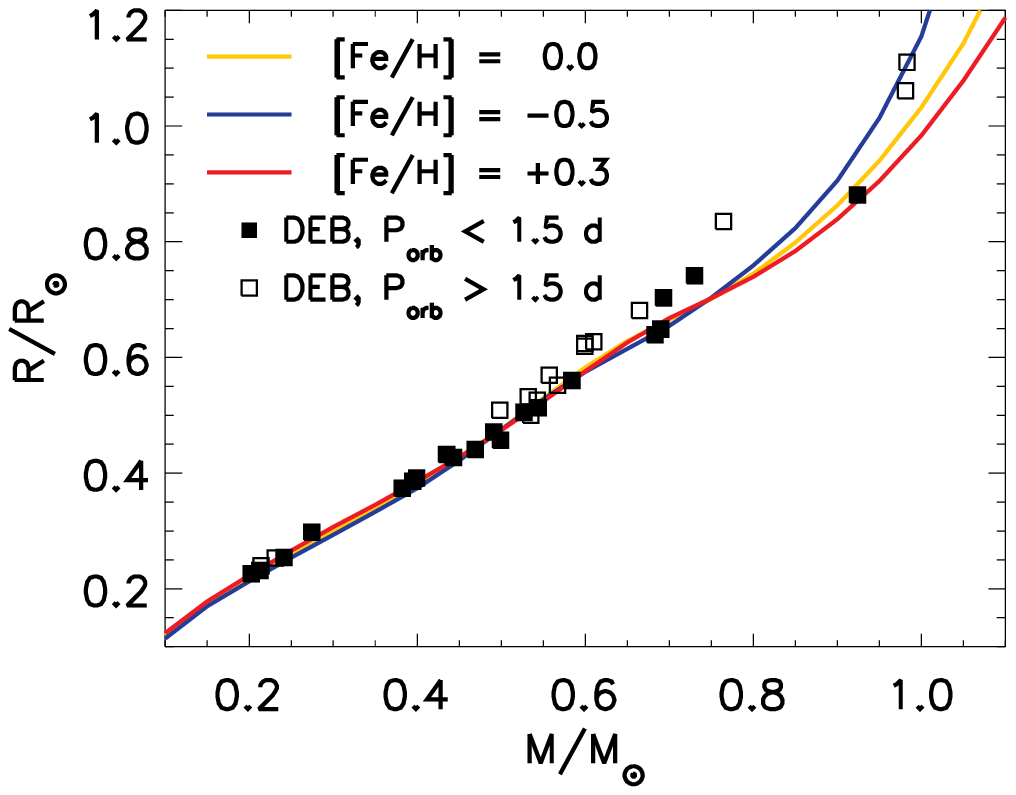}
\includegraphics[width=0.45\textwidth]{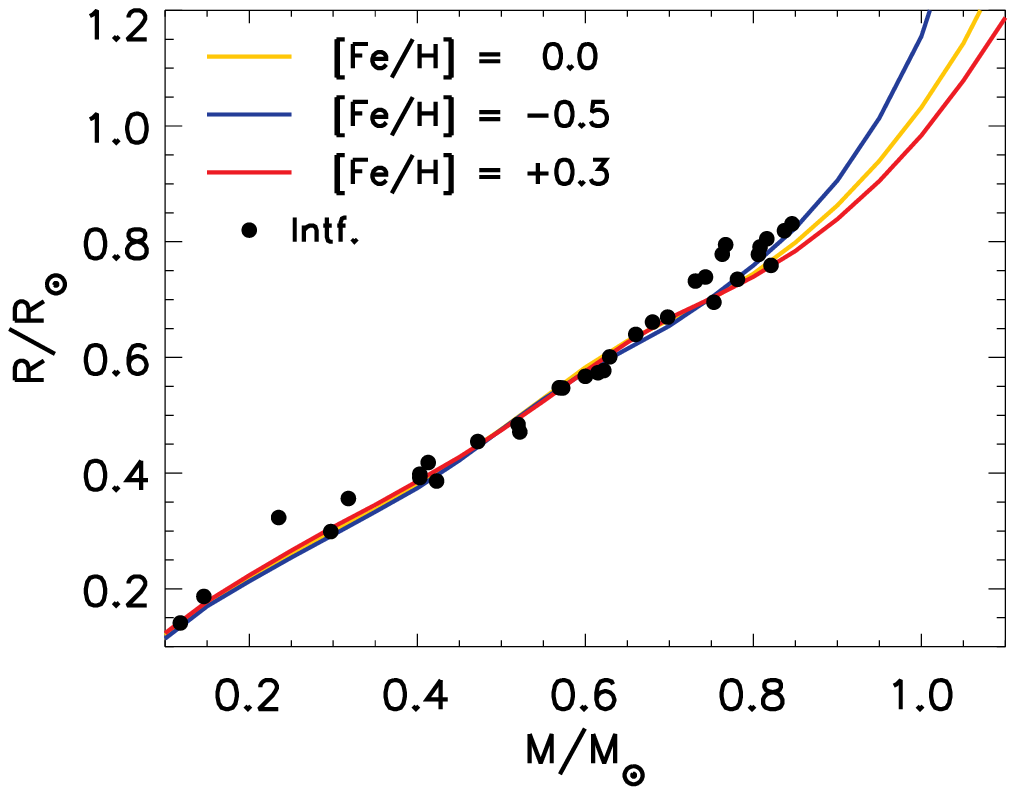}
\caption{Empirical and theoretical mass--radius relations. Upper panel: DEB sample; lower panel: interferometric sample; theoretical $5$ Gyr isochrones (with solar-calibrated $\alpha$, metallicity as shown) are also plotted for comparison. Short-period DEBs are shown as empty symbols.}
\label{fig:MRboth}
\end{center}
\end{figure}

\subsection{The $M$--$R$ relation}

A comparison between our theoretical mass--radius relationship and both data samples is shown in Figure~\ref{fig:MRboth}.
The isochrones plotted have $[{\rm Fe/H}]=-0.5,\,0.0,\,0.3$ and an age of $5$ Gyr.  
Note that the theoretical $M$--$R$ relation is insensitive to metallicity for $M\lesssim 0.7 M_\odot$.
A larger deviation from the theoretical isochrones is immediately apparent for the short-period DEBs (plotted as open squares in the Figure). 
In the interferometric sample, the largest disagreement is for $M \approx 0.4\, M_\odot$.

A quantitative measure of the discrepancy between observed radii and the model predictions can be defined as follows:
\begin{equation*}
\frac{\delta R}{R_{\rm obs}} = \frac{R_{\rm obs} - R_{\rm mod}}{R_{\rm obs}},
\end{equation*}
where $R_{\rm obs}$ is the observed radius and $R_{\rm mod}$ is its theoretical counterpart, derived from the models.

\subsubsection{The $M$--$R$ relation, DEB sample}

\begin{figure}
\begin{center}
\includegraphics[width=0.45\textwidth]{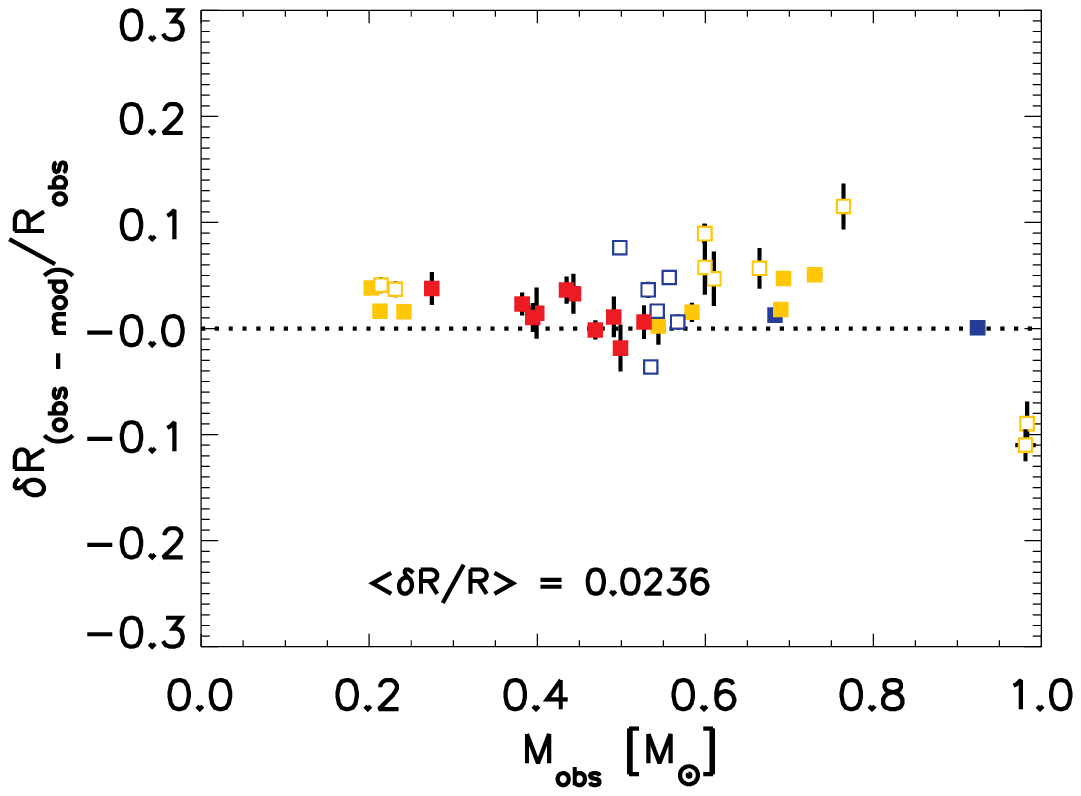}
\includegraphics[width=0.45\textwidth]{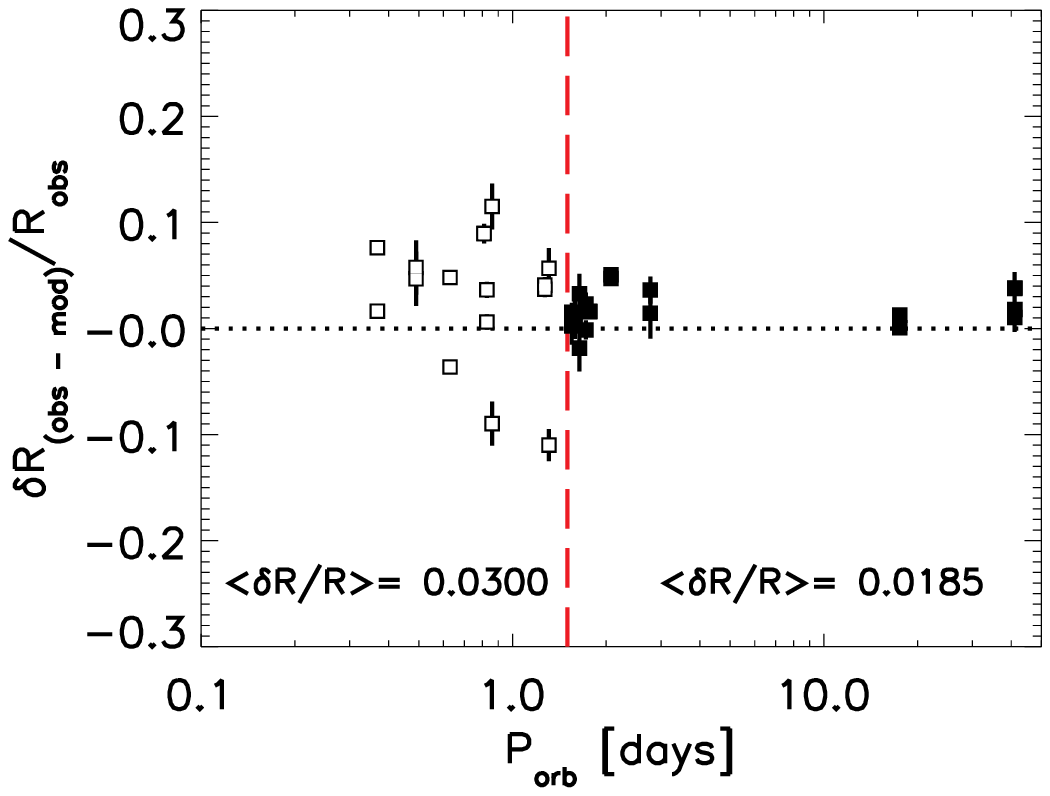}
\includegraphics[width=0.45\textwidth]{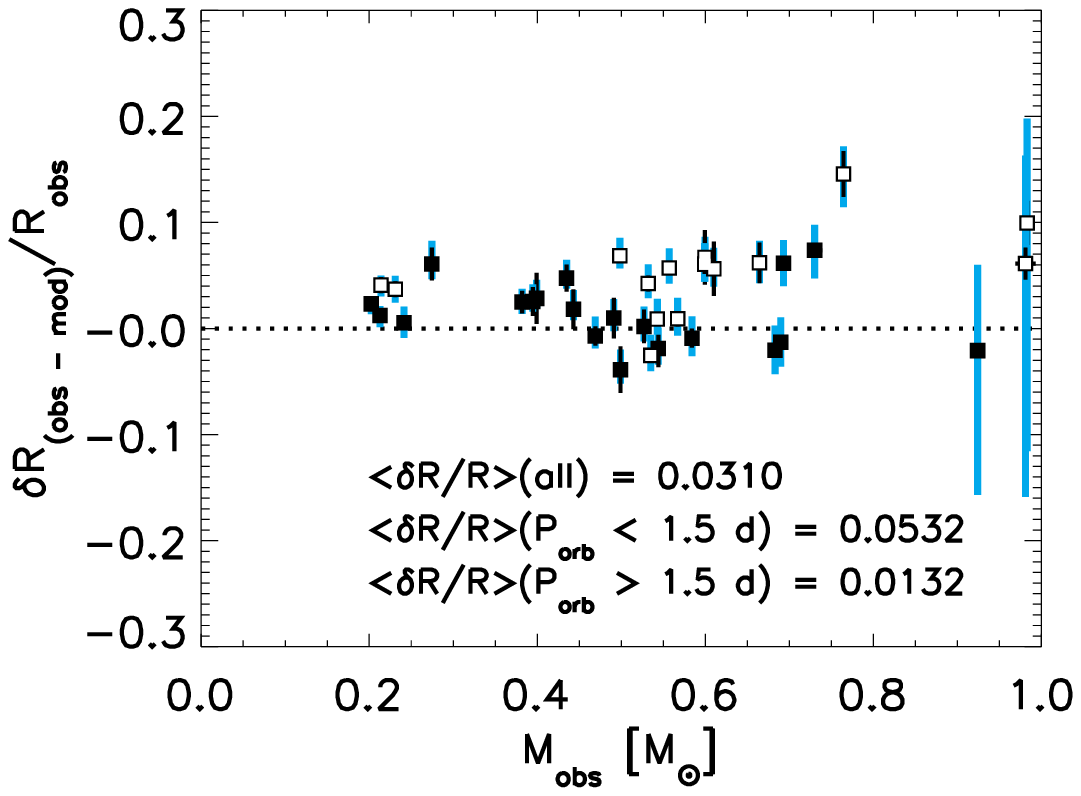}
\caption{Radius discrepancy in the DEB sample; $R_{\rm mod}$ interpolated in mass. 
Open (filled) squares represent the components of systems with orbital periods shorted (longer) than $1.5$ d.
Upper panel: radius discrepancy as a function of the observed mass, obtained from our isochrones with the same metallicities and ages determined by \citet{Feiden_Chaboyer:2012}. 
The color of the symbol encodes the metallicity assigned to the star: metal-poor, $[{\rm Fe/H}] < -0.25$: blue; solar, $-0.25\leq[{\rm Fe/H}]\leq +0.15$: yellow; metal-rich, $[{\rm Fe/H}] > +0.15$: red.
Middle panel: as in the upper panel, but as a function of the orbital period. 
Lower panel: radius discrepancy as a function of observed mass, but using the same theoretical isochrones for all the stars in the sample, i.e. with fixed metallicity (solar) and age ($5\pm 4$ Gyr). The blue bars span the age range.}
\label{fig:mrdebs}
\end{center}
\end{figure}

The tightest observational constraint on the mass--radius relationship is provided by the DEBs, due to the high accuracy in the determination of {\it both} the mass {\it and} the radius.
The discrepancy $\delta R/R_{\rm obs}$ for the DEB sample is plotted in Figure~\ref{fig:mrdebs}. 

In all three panels of the Figure, the value of $R_{\rm mod}$ for each star in the DEBs sample has been determined by interpolating our theoretical $M$--$R$ isochrones with the observed mass of the star.  

In the upper panel, we used theoretical isochrones of the same age and metallicity reported to be the best-fitting by \citet{Feiden_Chaboyer:2012}.
We obtain an average $\delta R/R$ of $2.4\%$, consistent with the findings of these authors.
No correlation seems to be present between $\delta R$ and the metallicity of the star (represented by the color of the symbols, see the caption of the Figure).
Interestingly, systems with orbital periods shorter than $1.5$ days (open symbols) are significantly more discrepant than the others, as can be seen in the middle panel of Figure~\ref{fig:mrdebs}, where $\delta R/R_{\rm obs}$ is plotted as a function of $P_{\rm orb}$ (see also the average $\delta R/R_{\rm obs}$ for these two subsamples, shown in the legend).
This was already found by \citet{Kraus_ea:2011}, although these authors caution that this effect could be a spurious result of the light curve analysis.

The availability of model-independent constraints on either the age or the metallicity of DEBs systems is very scarce, leaving these two parameters essentially freely adjustable (with a few exceptions, e.g. KOI-126, \citealt{Carter_ea:2011}).
To investigate the impact of the assumed age and metallicity on the radius discrepancy, we have recalculated the $\delta R$ using values of the theoretical radii $R_{\rm mod}$ interpolated from a fixed solar metallicity, fixed $5$ Gyr $M$--$R$ isochrone (lower panel of Figure~\ref{fig:mrdebs}).
The resulting average $\delta R/R_{\rm obs}$ is about $3\%$, still comparable with the value found by \citet{Feiden_Chaboyer:2012}.
Varying the age within the range $1$--$9$ Gyr, on the other hand, leads to ``age error bars" (shown in blue in the Figure) of the same size of or smaller than the observational uncertainty for all but three stars in the sample (which have $M\gtrsim 0.9 \, M_\odot$), thus not affecting significantly the overall conclusions.
Note that the correlation between discrepancy and short orbital periods is robust to these different choices of age and metallicity (see the legend in the lower panel of Figure~\ref{fig:mrdebs}).

\begin{figure}
\begin{center}
\includegraphics[width=0.45\textwidth]{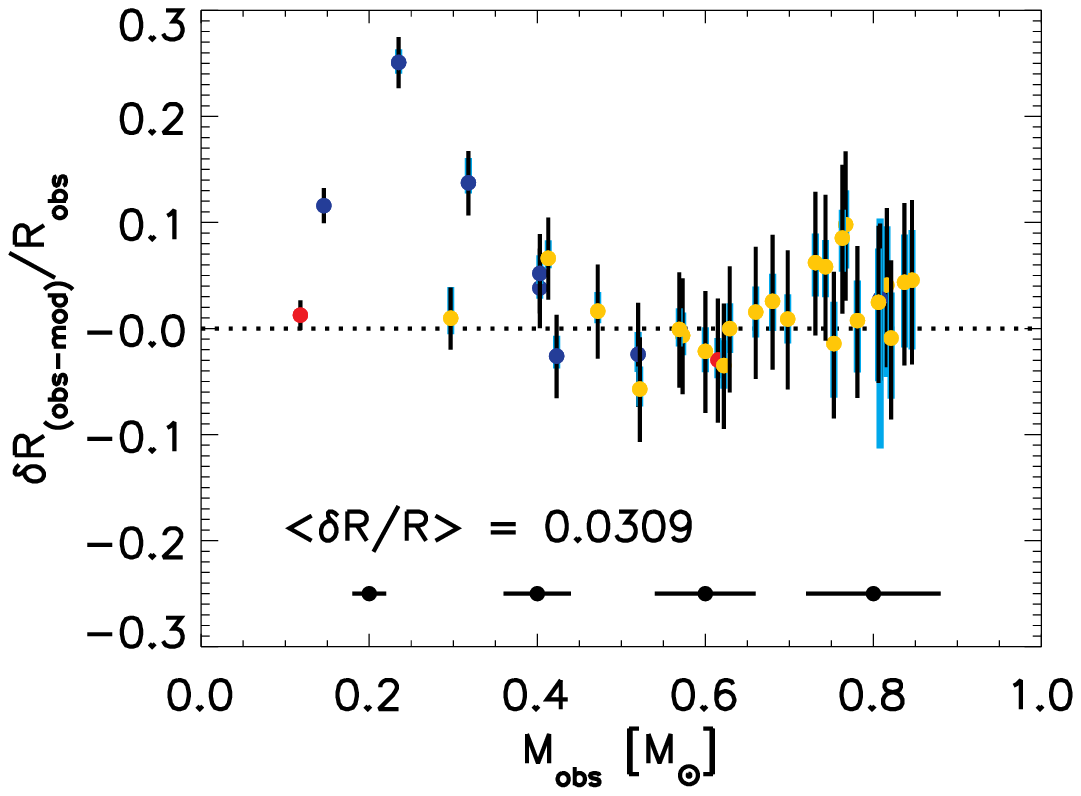}
\includegraphics[width=0.45\textwidth]{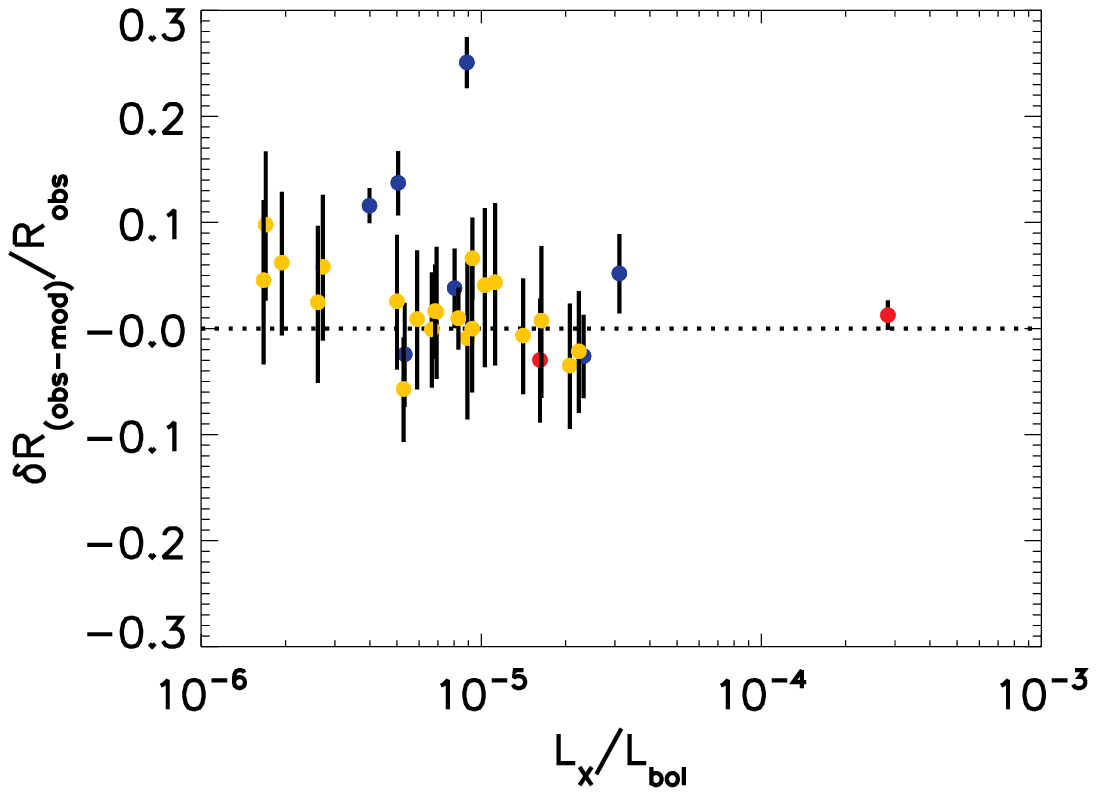}
\caption{Radius discrepancy in the interferometric sample; $R_{\rm mod}$ interpolated in mass. Upper panel: radius discrepancy as a function of observed mass (mass error bars shown at the bottom of the panel, $y$-error bars fixed to $10\%$.). An age of $5$ Gyr was assumed for all the stars; the effect of a $\pm 4$ Gyr uncertainty is shown as a blue bar. The color of the symbols encodes the metallicity of the star: metal-poor ($[{\rm Fe/H}] < -0.25$): blue; solar ($-0.25\leq[{\rm Fe/H}]\leq +0.15$): yellow; metal-rich ($[{\rm Fe/H}] > +0.15$): red. Lowe panel: as in the upper panel, but as a function of the activity indicator $L_X/L_{\rm bol}$.}
\label{fig:mrintf}
\end{center}
\end{figure}

\subsubsection{The $M$--$R$ relation, inteferometric sample}

A similar analysis was performed for the interferometric sample. 
Values of $R_{\rm mod}$ for each star have been calculated by interpolating in mass from isochrones of $5\pm4$ Gyr (the same for the whole sample) and metallicities equal to those quoted by \citet{Boyajian_ea:2012b}. 
The results are shown in Figure~\ref{fig:mrintf}; blue bars represent the uncertainty arising from the age range, while the colors of the symbols encode the metallicity information, as before.
It should be emphasized that the comparison with the results on the $M$-$R$ relationship for the DEBs sample, discussed previously, is hindered by the much larger errors on $R_{\rm mod}$, which are in turn due to the larger errors in the masses derived from a mass--magnitude relation.

For the interferometric sample, the average $\delta R/R_{\rm mod}\simeq 3\%$ is compatible with that of the DEBs sample.
Unfortunately, for the majority of the stars, this value is comparable with the uncertainty in $R_{\rm mod}$. 
A notable exception are the eight stars of $M\lesssim 0.42 \; M_\odot$ in the sample, which appear to have inflated radii to the level of several sigma \citep[see also][]{Boyajian_ea:2012b}. 
Note that five of these eight stars are reported to have sub-solar metallicity. 
We checked whether an erroneous metallicity determination could explain the anomalous behaviour of these stars: using a solar value of $Z$ for the theoretical isochrone, their discrepancy is lessened, but it remains significant. 

Our analysis so far suggests that the empirical $M$--$R$ relations of both the DEBs and the interferometric samples are, on the average, in agreement with their theoretical counterpart within $2$--$3\%$. 
Higher values of the discrepancy exist, but they are more the exception than the rule.
This conclusion appears to be rather insensitive to the choice of the (poorly constrained) age and metallicity.
However, the eight lowest mass stars in the interferometric sample challenge this claim, with an average discrepancy of the order of $8.5\%$.
This result is robust to changes in the age or the metallicity of these stars. 
Notably, no such a trend seems to be present at the low mass end of the DEBs sample.

\begin{figure}
\begin{center}
\includegraphics[width=0.45\textwidth]{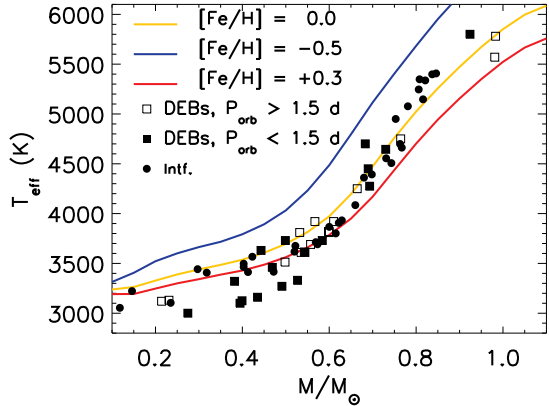}
\caption{Mass--effective temperature relation for the DEBs (squares) and the interferometric sample (circles), compared with theoretical isochrones of various metallicity. The error on the mass of single stars is of $10\%$, while the effective temperatures of DEBs have errors of about $100$ K; other error bars are within the size of the symbols.}
\label{fig:MRTeff}
\end{center}
\end{figure}

\subsection{The $M$--$T_{\rm eff}$ relation}

Radius deviations are usually observed to be associated with a concurrent effective temperature deviation, in such proportions to keep the luminosity constant \citep[e.g.][]{Torres_ea:2010}.
It is therefore interesting to compare the mass--radius relation with the mass--$T_{\rm eff}$ relation.
Effective temperature measurements for DEBs typically have large errors, of the order of $100$ K.
Nevertheless, we compare the theoretical and empirical $M$--$R$ and $M$--$T_{\rm eff}$ relations, for both the DEBs and the interferometric samples, in Figure~\ref{fig:MRTeff}.

As we have already seen, the mass--radius relation of the two samples are, in general, compatible with each other.
The largest deviations from our theoretical isochrones are observed in the mass ranges $0.6$--$0.8\; M_\odot$ (both samples) and below about $0.4\; M_\odot$ (most prominently in the interferometric sample).  
Figure~\ref{fig:MRTeff} shows, on the other hand, that the DEBs have significantly cooler temperatures than the models for masses $\lesssim 0.55 \; M_\odot$, while such a trend is much less evident among the single stars.

It is worth noting that \citet{Boyajian_ea:2012b} reported a systematic difference between the effective temperature of stars in their sample and those from DEBs. 
The same authors have also stressed that the DEBs measurements could be affected by systematic errors and careful validation of these data is required before any conclusion can be drawn from them.

\begin{figure}
\begin{center}
\includegraphics[width=0.45\textwidth]{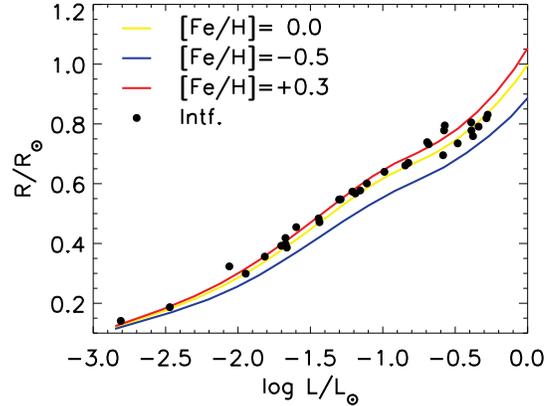}
\includegraphics[width=0.45\textwidth]{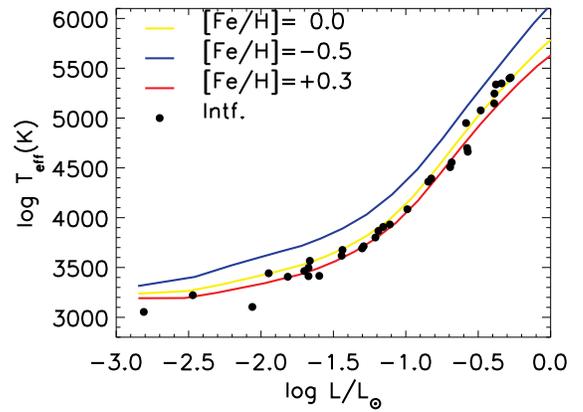}
\caption{Luminosity--radius (upper panel) and luminosity--effective temperature (lower panel) relations for stars in the interferometric sample, compared with $5$ Gyr isochrones of various metallicity.}
\label{fig:LRTeff}
\end{center}
\end{figure}

\subsection{The $L$--$R$--$T_{\rm eff}$ relations}

The high precision in the determination of radius, luminosity, and effective temperature for the interferometric sample allows us to test the theoretical $L$--$R$--$T_{\rm eff}$ relations as well.

Figure~\ref{fig:LRTeff} shows the comparison between our $5$ Gyr isochrones (calculated with $[{\rm Fe/H}] = 0.0,\, -0.5$, and $+0.3$ and solar-calibrated $\alpha$) and the interferometric data in the $L$--$R$ and $L$--$T_{\rm eff}$ planes.
Note that the models display a prominent metallicity dependence which was absent in the mass--radius relation (Figure~\ref{fig:MRboth}).
Remarkably, even if the range of the measured $[{\rm Fe/H}]$ is comparable to that of the theoretical isochrones plotted, the data are less scattered, in both planes.

As was done in the previous Section, we define the radius and effective temperature deviations as:
\begin{equation*}
\frac{\delta R}{R_{\rm obs}} = \frac{R_{\rm obs} - R_{\rm mod}}{R_{\rm obs}}\ \ ;
\ \ \ \
\frac{\delta T}{T_{\rm obs}} = \frac{T_{\rm obs} - T_{\rm mod}}{T_{\rm obs}},
\end{equation*}
where $R_{\rm mod}$ and $T_{\rm mod}$ are calculated interpolating in luminosity (with the observed value $L_{\rm obs}$) the $L$--$R$ and $L$--$T_{\rm eff}$ theoretical relations, respectively.

\begin{figure}
\begin{center}
\includegraphics[width=0.45\textwidth]{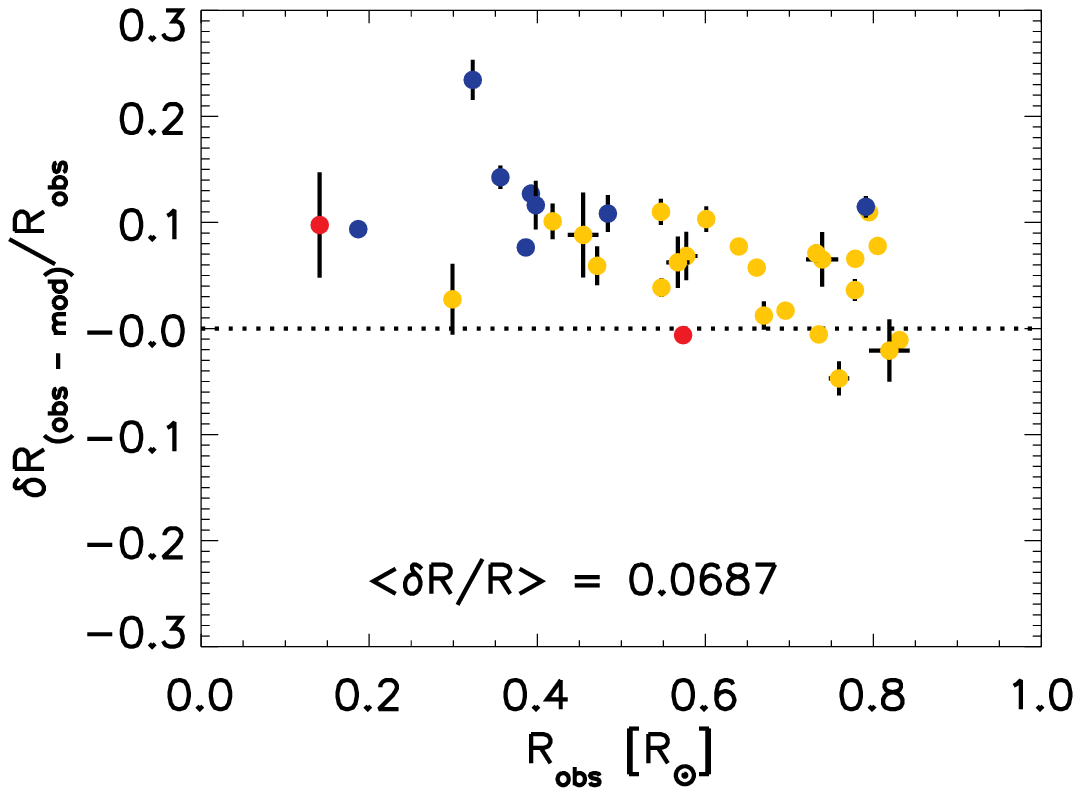}
\includegraphics[width=0.45\textwidth]{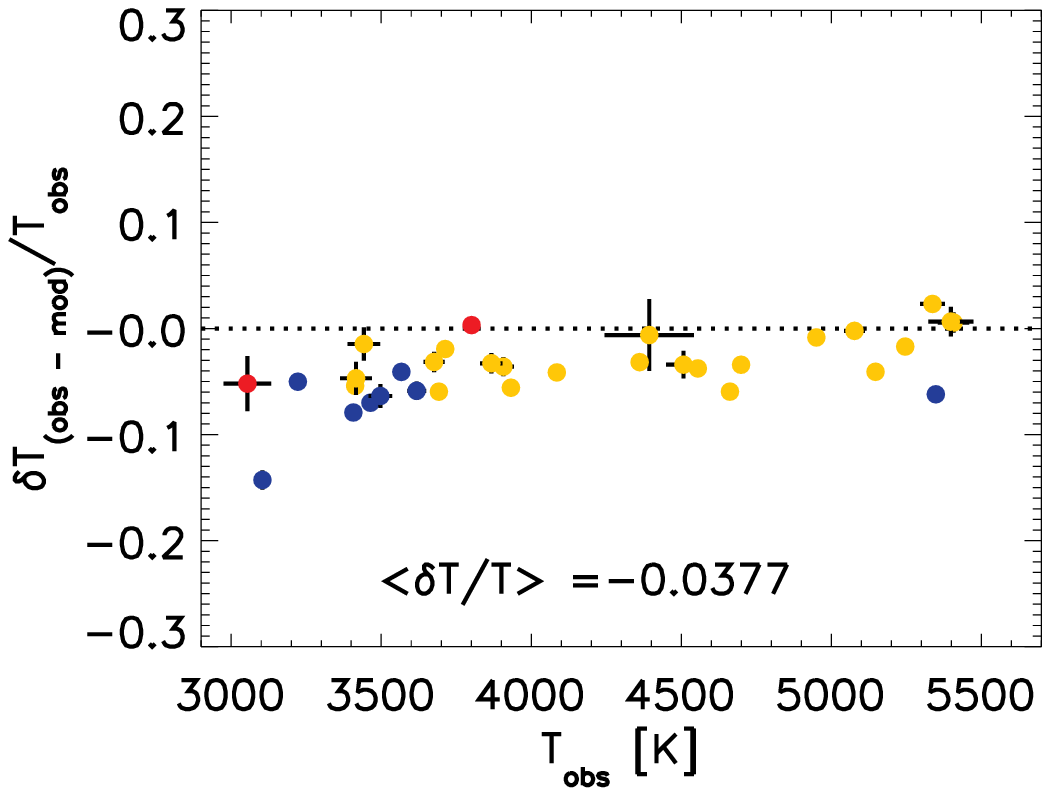}
\caption{Radius (above) and effective temperature (below) discrepancies in the interferometric sample; $R_{\rm mod}$ is interpolated in luminosity. A fixed age of $5$ Gyr was assumed for all stars. The color of the symbols represents the stellar metallicity (blue: metal-poor, yellow:solar, red: metal-rich).} 
\label{fig:lrteff_diff}
\end{center}
\end{figure}

For each star in the sample, $5$ Gyr isochrones with the observed metallicity were used in the interpolation; the results are shown in Figure~\ref{fig:lrteff_diff}. 
A significant trend emerges of the observed radii being inflated (by about $7\%$, on average) and the effective temperatures being cooler (by about $4\%$) than predicted by theoretical models. 
The radius and effective temperature deviations are significant for $R\lesssim 0.7 \; R_\odot$ and $T_{\rm eff} \lesssim 5000$ K.
The $R$ and $T_{\rm eff}$ discrepancies are not correlated with the stellar activity (as measured by the X luminosity proxy $L_X/L_{\rm bol}$).
These results are consistent with those of \citet{Boyajian_ea:2012b}. 

\begin{figure}
\begin{center}
\includegraphics[width=0.45\textwidth]{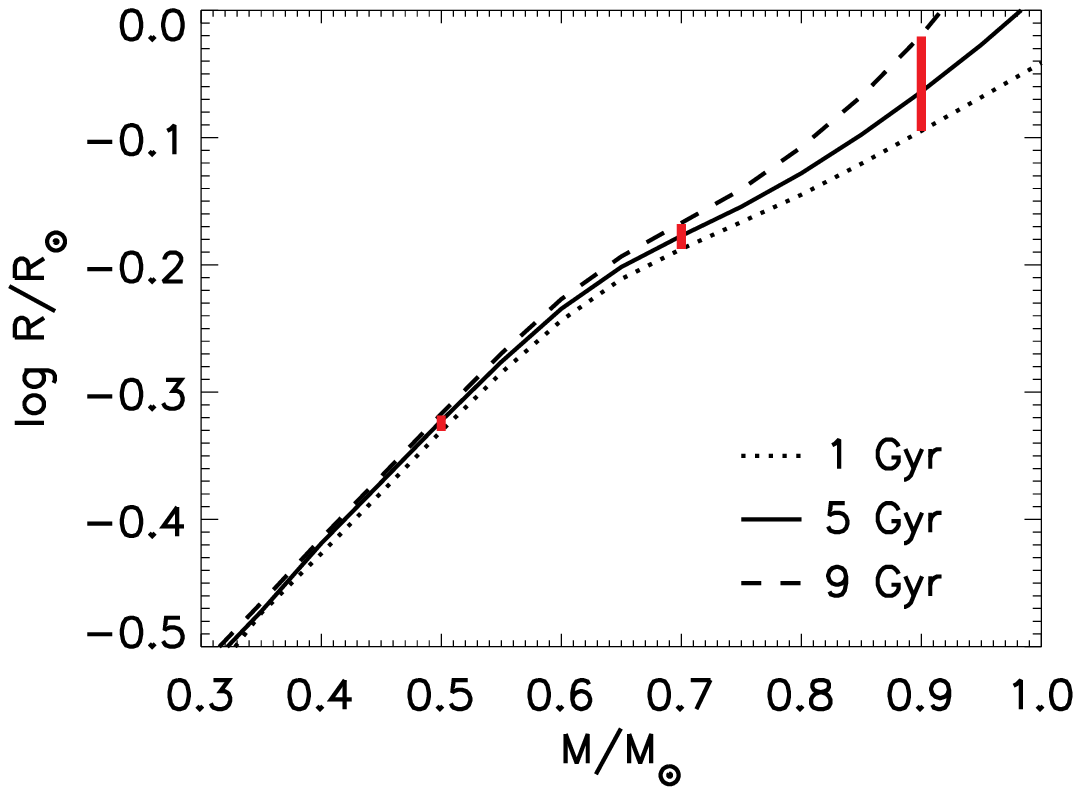}
\includegraphics[width=0.45\textwidth]{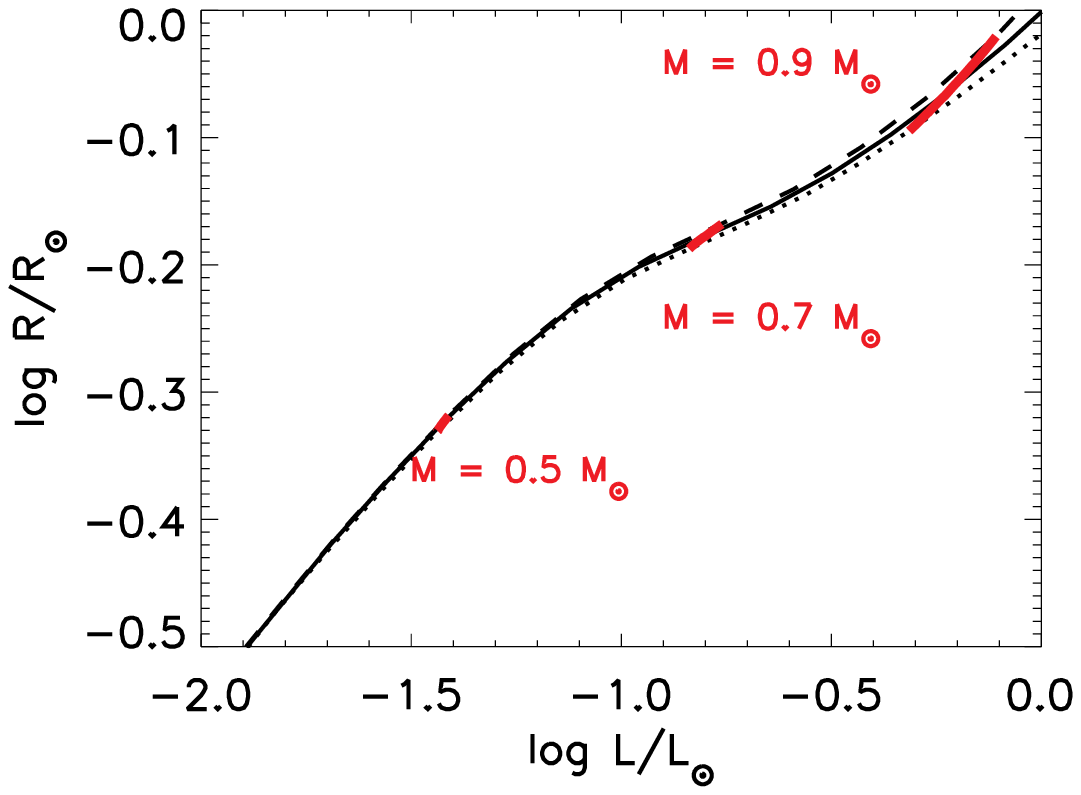}
\caption{Different age effect in the mass--radius (above) and luminosity--radius (below) planes.  The dotted, solid, and dashed lines are the $1$, $5$, and $9$ Gyr isochrones, respectively; the red bars show the evolutionary tracks of stars of $0.5$, $0.7$, and $0.9\; M_\odot$.}
\label{fig:MvsL}
\end{center}
\end{figure}

Clearly, for the same interferometric sample, we obtain quite a different picture for the radius discrepancy depending on whether the theoretical radius $R_{\rm mod}$ is calculated by interpolating in mass or in luminosity.
In particular, the average discrepancy $\left<\delta R/R\right>$ is more than a factor of two larger in the latter case.
This is, however, not surprising, since the theoretical mass--radius and mass--luminosity relations have very different properties from a stellar evolution standpoint.
First of all, the $L$--$R$ and $L$--$T_{\rm eff}$ relations have a much stronger dependence on metallicity than the $M$--$R$ relation (see Figure~\ref{fig:LRTeff} and the lower panel of Figure~\ref{fig:MRboth}, respectively).
Moreover, the uncertainty in age plays a very different role for large enough masses, as is illustrated in Figure~\ref{fig:MvsL}.
In the mass--radius plane, a star moves along a vertical track as it ages;  this effect was represented by the blue ``age bars" in Figures~\ref{fig:mrdebs}-\ref{fig:mrintf}.
In the luminosity--radius plane, on the contrary, stars move approximately along the isochrone as they age. 
This can lead to compensation effects between the age and the mechanism responsible of the radius discrepancy (whatever it may be), which cannot be disentangled without independent information on the stellar age.
The age shift along the isochrone starts to be significant above $0.7\; M_\odot$, and it is very likely to be the cause of the threshold for non-significant $\delta R$ and $\delta T_{\rm eff}$, shown in Figure~\ref{fig:lrteff_diff}.

As has been suggested by \citet{Boyajian_ea:2012b}, the $R$ and $T_{\rm eff}$ deviations themselves could be explained if the effect of metallicity on the stellar parameters is exaggerated by theoretical models.
This would lead to a larger discrepancy when interpolating in luminosity than in mass, while the residual, more modest, discrepancy in the $M$--$R$ relation could be due to an entirely different mechanism. 

\subsection{Effect of the MLT parameter variation}

The MLT parameter $\alpha$ is a  measure of the efficiency of convection. 
In models of solar-like stars, the global radius is quite sensitive to the value of $\alpha$, its reduction resulting in larger radii \citep[see][]{JCD97}. 
In the case of fully convective stars, the impact of $\alpha$ on the radius is much more modest \citep{BCAH98}.
Altough the standard practice in stellar modelling is to calibrate $\alpha$ on the Sun, lower effective values of this parameter have been used in the past as a means to phenomenologically account for the additional physics (e.g. magnetic fields) responsible for the radius discrepancy \citep[e.g.][]{Chabrier_ea:2007}.

\begin{figure}
\begin{center}
\includegraphics[width=0.45\textwidth]{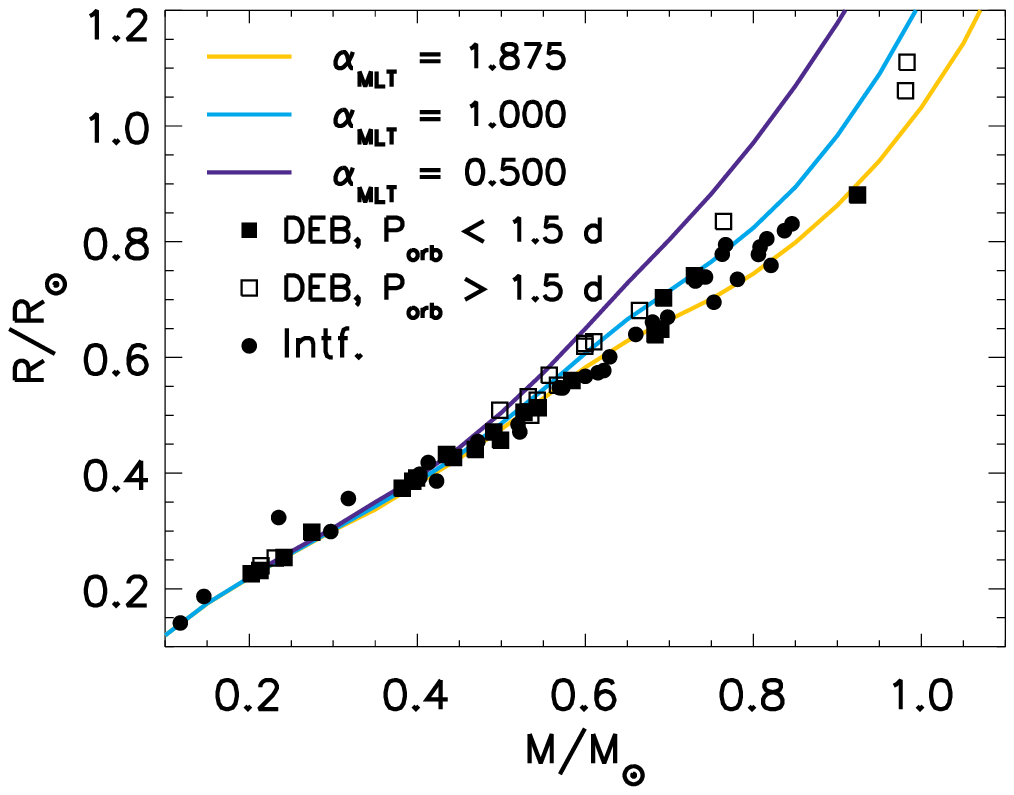}
\includegraphics[width=0.45\textwidth]{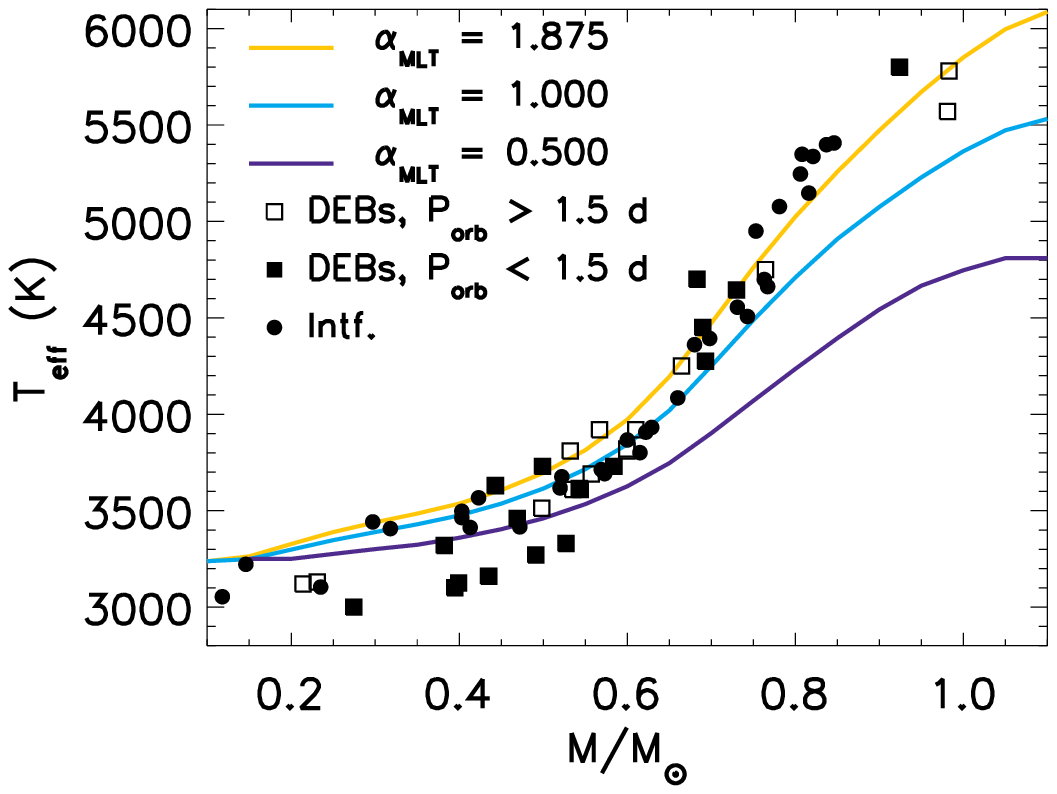}
\caption{Impact of the MLT $\alpha$ parameter on the mass--radius (above) and mass--effective temperature relations (below). The $5$ Gyr isochrones shown have solar metallicity and $\alpha$ as detailed in the legend. Data from both the DEB and the interferometric samples are also plotted.}
\label{fig:varalpha}
\end{center}
\end{figure}

We compare the mass--radius and mass--$T_{\rm eff}$ isochrones for various values of $\alpha$ with the data from the DEB and the interferometric samples in Figure~\ref{fig:varalpha}.
As $\alpha$ is freely adjusted in this comparison, we restrict ourselves to qualitative remarks, without calculating the $\delta R$ and $\delta T_{\rm eff}$.
Figure~\ref{fig:varalpha} is consistent with the standard results discussed above (larger radii for reduced-$\alpha$ models of solar-like stars, very weak sensitivity for fully convective models).
It also shows that the data are compatible with a moderately reduced values of the MLT parameter, i.e. $1\lesssim \alpha \lesssim 1.875$, while $\alpha = 0.5$ seems to be too extreme.

\section{Discussion and Conclusions}

We have calculated models of low mass stars in the range $0.1$--$1.25\; M_\odot$ for various choices of the composition and of the MLT parameter $\alpha$.
In our calculations, we have used an improved version of the YREC stellar code, optimized to provide a realistic treatment of the EOS and atmospheric boundary conditions for low mass stars ($M\lesssim 0.6\; M_\odot$).
The present models complement and update the low mass end of the Y$^2$ isochrones \citep{Yi_ea:2001,Demarque_ea:2004}.

We have compared the theoretical relations among fundamental stellar parameters predicted by our models with the best currently available data from detached eclipsing binaries and single stars interferometric measurements.
We find that radius deviations are present, on the average, at the $\approx 3\%$ level in both samples when the theoretical isochrones are interpolated in mass, consistently with the results of \citet{Feiden_Chaboyer:2012}. 
Among the binaries, the most discrepant stars are those members of short orbital period systems. 
Among the single stars, very large deviations ($10$--$20\%$) are found for stars of mass $\lesssim 0.4 \; M_\odot$.
This very discrepant subgroup has no counterpart in the DEB sample; notably, it is composed of predominantly metal-poor stars.
No correlation is observed with the activity indicator $L_X/L_{\rm bol}$.
The values adopted for the metallicity and the age of the stars do not affect significantly these results, since the theoretical mass--radius relation is almost insensitive to metallicity.

These results are consistent with the commonly accepted explanation of the $R$ and $T_{\rm eff}$ discrepancies as manifestations of enhanced magnetic activity \citep[see, e.g.][]{Chabrier_ea:2007}.
If a binary system is old enough (e.g., age $\gtrsim 1$ Gyr for $P_{\rm orb} \leq 5$ d, see figure 9 of \citealt{Meibom_Mathieu:2005}), its components will have synchronized their rotation periods with the orbital period via tidal interaction. 
Close-in DEB systems are thus expected to be more magnetically active due to the faster rotation rate powering their dynamo \citep{Charbonneau:2013}.
The interferometric stars, on the other hand, are single and likely old enough ($\approx 5$ Gyr) that their level of activity has already declined \citep[e.g.,][]{Wright_ea:2011}. 
Theoretical efforts are under way to incorporate the effects of magnetic fields in 1D stellar models with a self-consistent approach (based on the prescription of \citealt{Lydon_Sofia:1995}; see also \citealt{Feiden_Chaboyer:2012b}). 
We plan to tackle this intriguing problem in a forthcoming paper.

The very precise determinations of luminosity and effective temperature, along with radius, available for the interferometric sample, allow us to perform a complementary analysis on these stars, to test the $L$--$R$--$T_{\rm eff}$ relations. 
In this analysis, the luminosity is held constant, as is (approximately) expected to be the case even for discrepant stars, where the radius and effective temperature deviations should compensate each other \citep{Torres_ea:2010}.
We obtain much larger discrepancies in this case (of the order of $7\%$ and $4\%$, on the average, for radii and effective temperatures, respectively).
A trend of vanishing deviations above $0.7\; M_\odot$ is also found. 
However, rather than a feature of the underlying physical mechanism causing the radius discrepancy, we interpret the lack of observed deviation for higher mass stars as the consequence of compensation effects due to the age scatter, since evolutionary tracks are almost parallel to the theoretical isochrones in the $L$--$R$ plane.
For stars of mass below the threshold of $0.7\; M_\odot$, the larger discrepancies suggest that the effect of metallicity on the global stellar parameters is overestimated by the theoretical models in comparison with what is observed in the data \citep[see also][]{Boyajian_ea:2012b}.

In DEBs, the existence of the radius discrepancy phenomenon seems to be established beyond doubt and its extent and physical origin to be understood at least at the most basic level. 
A major source of concern for this picture, however, is the possible presence of unaccounted systematic errors. 
\citet{Morales_ea:2010} have suggested that systematic errors in the radii determined from the light curve analysis of DEBs could arise due to the variable star spots distribution and size.  
Systematic errors of the order of $3\%$ would be sufficient to completely mask the signal (if any) of the genuine radius discrepancy \citep{Feiden_Chaboyer:2012}.

Stellar radii measured by interferometry, with precision comparable to DEBs, are very valuable to ascertain whether the radius discrepancy affects in the same way single and binary stars. 
Currently, the main differences between these two groups are in the $M$--$T_{\rm eff}$ relations and in the mild metallicity dependence at the low mass end of the $M$--$R$ relation of single stars, which is absent in binaries.
However, both effective temperatures of DEBs and metallicity determinations (in both samples, especially for M dwarfs) are notoriously difficult measurements and are plagued by large errors ($\approx 100$ K and $\approx 0.2$ dex, respectively).
More observational evidence needs to be gathered before attempting to interpret these differences.

\section*{Acknowledgements}
FS acknowledges support from the Leibniz Institute for Astrophysics Potsdam (AIP) through the Karl Schwarzschild Postdoctoral Fellowship.
FS is grateful to Dr. Sydney Barnes for reading the manuscript and providing useful comments, and to Prof. Alessandro Bressan for sending the latest version of the evolutionary tracks of the Padova group.
PD thanks Dr. Tabetha Boyajian for interesting discussion. 
\appendix

\section{Rotational parameters of the models}
\label{app:rotevol}

One of our aims in this work is to provide a grid of models suitable for studies of rotational evolution of low mass stars, their magnetic activity, and the properties of their convective envelopes, which are believed to be the seat of the stellar dynamo \citep[see, e.g.,][]{Charbonneau:2013}.
We thus include in our evolutionary tracks rotation-related parameters, such as moments of inertia and global and local convective turnover timescales.

\begin{figure}[htdp]
\begin{center}
\includegraphics[width=0.45\textwidth]{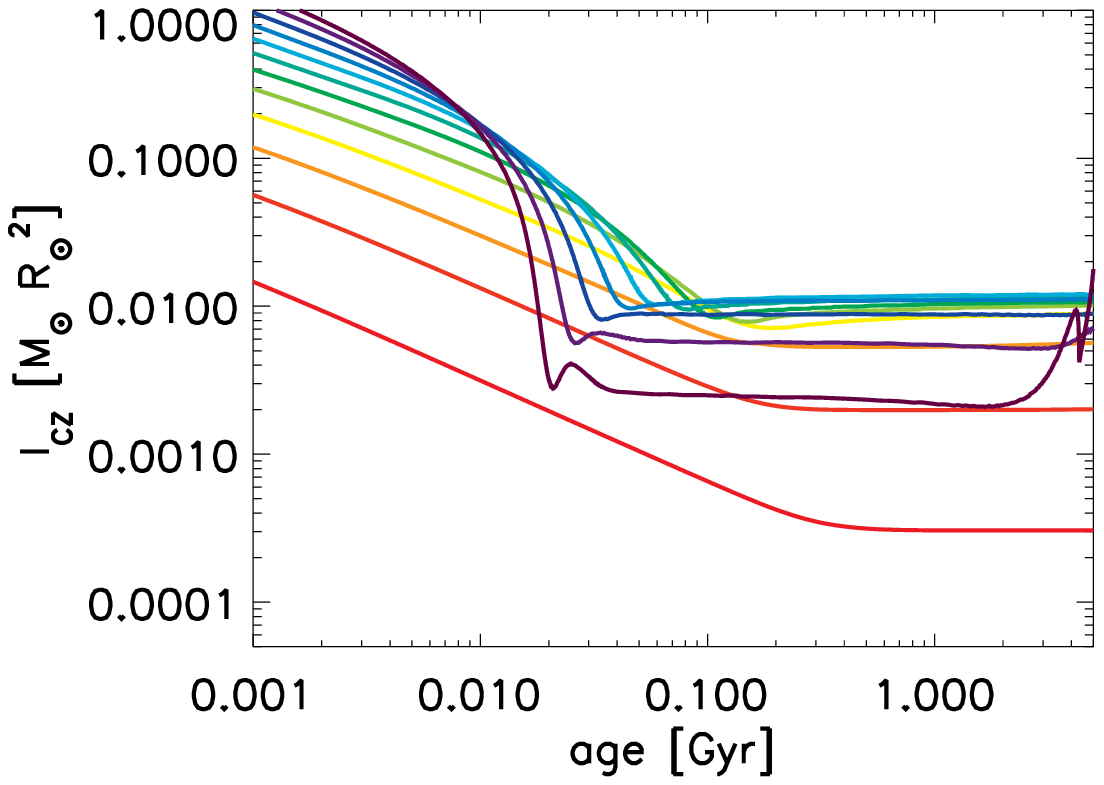}
\includegraphics[width=0.45\textwidth]{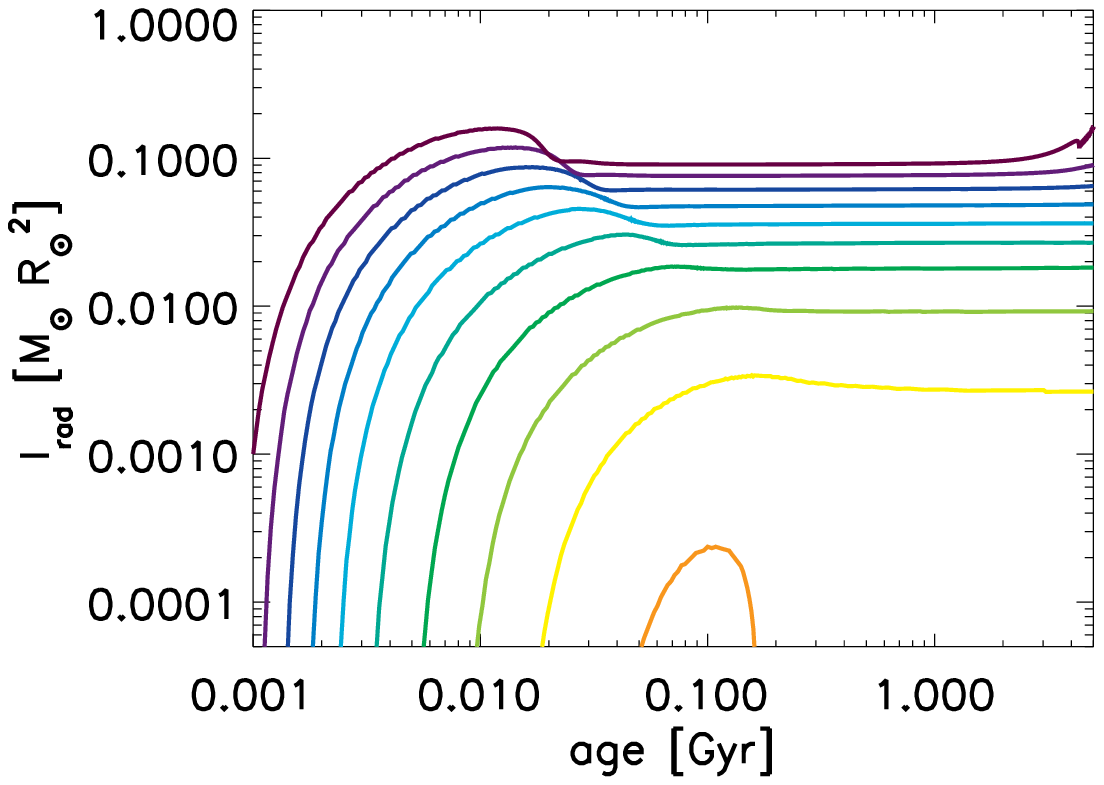}
\caption{Moments of inertia of the convective envelope (left panel) and of the radiative core, if present (right panel) for our solar metallicity, solar-calibrated $\alpha$ models with $M=0.1, 0.2, \dots 1.2$ $M_\odot$ (from red to violet).}
\label{fig:momi}
\end{center}
\end{figure}

For illustrative purposes, we show in Figure~\ref{fig:momi} the moments of inertia of the convective envelope and of the radiative zone (if present) for our solar metallicity models.

The convective turnover time scale $\tau_{\rm c}$ represents a characteristic time for the rise of a convective element through the stellar convection zone.
Following \citet{Kim_Demarque:1996}, we have calculated a ``global" and a ``local" estimate of $\tau_{\rm conv}$:
\begin{eqnarray*}
\tau_{\rm c,global} &=& \int_{R_{\rm BCZ}}^{R_*} \frac{dr}{v_{\rm conv}}; 
\\
\tau_{\rm c,local} &=& \frac{\ell^*}{v_{\rm conv}^*};
\end{eqnarray*}
the local definition refers to a distance of one half of the mixing length above the bottom of the convective envelope, i.e., $\ell^*=\alpha H_P/2$ and $v_{\rm conv}^* = v_{\rm conv}(R_{\rm BCZ}+\ell^*)$.
 
\begin{figure*}[htdp]
\begin{center}
\includegraphics[width=0.9\textwidth]{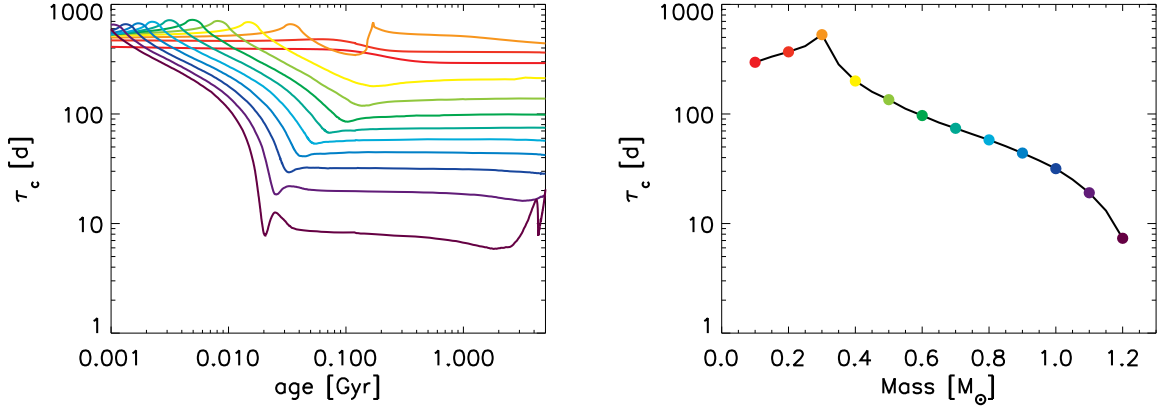}
\caption{Left: time evolution of the convective turnover timescales, according to the global definition given in the text. Right: convective turnover timescale as a function of mass at the age of $500$ Myr. In both panels, the models have solar metallicity, solar-calibrated $\alpha$, and $M=0.1, 0.2, \dots 1.2$ $M_\odot$ (from red to violet). }
\label{fig:tauc}
\end{center}
\end{figure*}

Even though the MLT is an admittedly crude approximation of real convection in the atmosphere \citep[e.g.][]{Kim_ea:1996, Ludwig_ea:1999,Tanner_ea:2013}, it is a remarkably reliable approximation of stellar convection in the deep layers \citep{Chan_Sofia:1989}.
The parameter $\tau_{\rm conv}$ has the advantage of providing a simple parametrization of how the properties of convection scale with, e.g., stellar mass.  
An example of that is the Rossby number, a dimensionless parameter calculated from $\tau_{\rm conv}$ and the surface rotation period:
\begin{equation*}
{\rm Ro} = \frac{P_{\rm rot}}{\tau_{\rm c}}.
\end{equation*}
Many studies on stellar activity have shown that the Rossby number is a critical parameter, marking the transition between different regimes in the chromospheric and X emission \citep{Noyes_ea:1984,Wright_ea:2011}.
Although our estimates of $\tau_{\rm c}$ are based on non-rotating models, this should not affect significantly the results when calculating $\rm Ro$, as the effects of rotation on the structure of low mass stars are quite moderate if the rotation rate is not extreme (i.e. far from break-up speed; see \citealt{Sills_ea:2000}).
The time evolution and mass dependence of $\tau_c$ for our solar composition subgrid are shown in Figure~\ref{fig:tauc}. 
Note that the value of $\tau_c$ for a star of given mass is approximately constant for the whole MS lifetime.

\begin{figure*}[htdp]
\begin{center}
\includegraphics[width=0.45\textwidth]{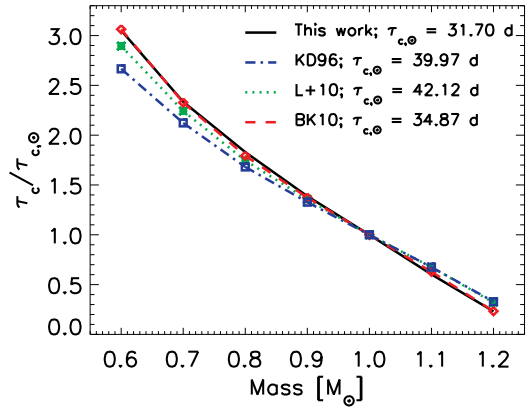}
\caption{Comparison between theoretical turnover time scales from this work and others in the literature. 
Values of $\tau_c$ at $500$ Myr for $0.6, 0.7,\dots, 1.2 \, M_\odot$; to enhance the comparison of the mass dependence predicted by the different sets of calculations, $\tau_c$ has been scaled over that of the $1\, M_\odot$ model (these values are shown for reference in the legend).}
\label{fig:tauc_compare}
\end{center}
\end{figure*}

Theoretical calculations of the convective turnover time scale have been reported by \citet{Kim_Demarque:1996}, \citet{Landin_ea:2010}, and \citet{Barnes_Kim:2010}.
The time evolution of $\tau_c$ calculated by these authors agrees quite well with our results (compare the left panel of Figure~\ref{fig:tauc} with, e.g., figure 3 of \citealt{Kim_Demarque:1996} or figure 3 of \citealt{Landin_ea:2010}).
When comparing the absolute values of $\tau_c$ at a given age, however, a $\approx 10\%$ scatter emerges.
Other choices of the input physics being equal, a strong dependence of $\tau_c$ on the value of the MLT parameter $\alpha$ can of course be expected.
To illustrate the influence of other parameters, we compare in Figure~\ref{fig:tauc_compare} the various sets of $\tau_c$ for $0.6, 0.7,\dots, 1.2 \, M_\odot$ at $0.5$ Gyr, scaling them over the value corresponding to the $1\, M_\odot$ model.
Once a constant (i.e., mass-independent) scale factor is accounted for, our calculations are in remarkably close agreement with those of \citet{Barnes_Kim:2010}.
For the other two sets of $\tau_c$, the scale factor depends on mass. 
This is most likely the consequence of different choices in the input physics to which stellar models have different sensitivity according to their mass, e.g., the treatment of the atmospheric boundary conditions.
It should also be noted that both the \citet{Kim_Demarque:1996} and the \citet{Landin_ea:2010} models take stellar rotation into account, which affects more strongly more massive models than less massive ones \citep{Sills_ea:2000}.

\begin{figure*}[htdp]
\begin{center}
\includegraphics[width=0.9\textwidth]{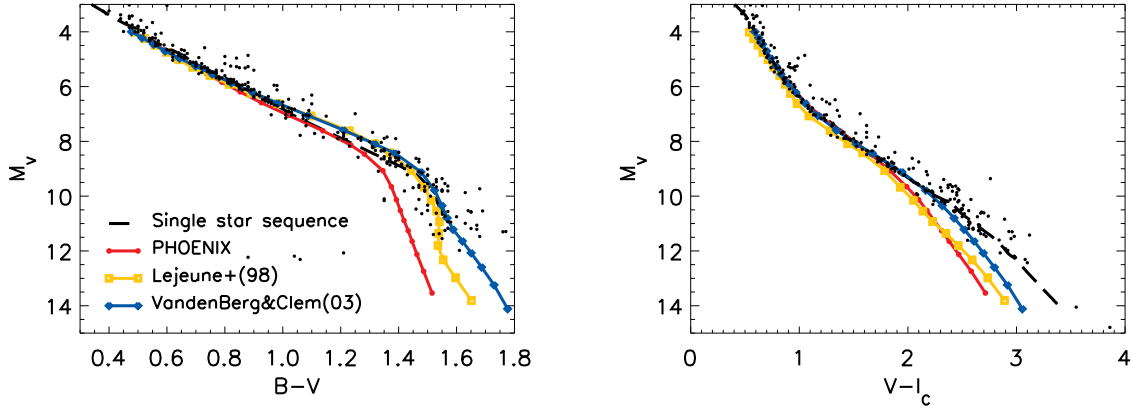}
\caption{CMDs for the Pleaides cluster and theoretical isochrones based on different choices of the color--effective temperature relations. The synthetic isochrones are calculated for solar metallicity, solar-calibrated $\alpha$, and an age of $100$ Myr. The single star Pleiades sequence from \citet{Stauffer_ea:2007} is also shown.}
\label{fig:colors}
\end{center}
\end{figure*}

\section{Color-$T_{\rm eff}$ transformations and synthetic color-magnitude diagrams}
\label{app:colteff}

To construct synthetic color-magnitude diagrams (CMDs), we need to transform from the theoretical variables $(\log g, T_{\rm eff})$ to colors.
Various such transformations are reported in the literature, either constructed purely from theoretical atmosphere models or established on the basis of semi-empirical calibrations.
We have compared the following color transformations:
\begin{itemize}
\item synthetic transformations derived from PHOENIX atmosphere models, available in various photometric systems from F.~Allard's web page\footnote{\texttt{http://perso.ens-lyon.fr/france.allard/}}. This is the natural choice for our models, ensuring consistency (whose importance has been stressed by \citealt{BCAH98}) with the surface boundary conditions used (see Sec.~\ref{sec:surfbc}); 
\item semi-empirical transformations in the $(UBV )_J(RI)_CJHKLL'M$  by \citet{Lejeune_ea:1998}, extending the grid produced by the same authors \citep{Lejeune_ea:1997} to include M dwarfs. Starting from a hybrid grid of theoretical stellar spectra from many literature sources, they developed a semi-empirical calibration of the spectral continua; color-$T_{\rm eff}$ relations were then calculated from the corrected spectra; 
\item semi-empirical transformation in the $(BV)_J(RI)_C$ system by \citet{VandenBerg_Clem:2003}, obtained from a grid of theoretical stellar spectra, applying empirical corrections based on constraints from clusters and field stars. 
\end{itemize}
All the color-$T_{\rm eff}$ relations above come in tabular form as a function of $(\log g,\, T_{\rm eff})$ for various metallicities; bolometric corrections and color indices are obtained through interpolation in these variables.

As an example, we compare our theoretical isochrones with data for the Pleiades in $(B-V)$ and $(V-I_C)$ CMDs in Figure \ref{fig:colors} . 
The data and the empirical single stars sequence are from the compilation by \citet{Stauffer_ea:2007}; the $BVI_C$ data are from various literature sources, among which the authors have favoured photoelectric photometry when possible (see references in the paper).
We adopt the same fundamental parameters for the cluster used by these authors, i.e., solar metallicity, a distance modulus $m-M = 5.62$ (corresponding to the ground-based distance estimate, i.e. $133$ pc), an age of $100$ Myr, and the following values of the reddening: $A_B=0.16$, $A_V=0.12$, $A_I=0.07$ \citep[see][]{Stauffer_ea:2007}.

The theoretical isochrones shown in the Figure were constructed from our solar metallicity models by applying the PHOENIX, \citet{Lejeune_ea:1998}, and \citet{VandenBerg_Clem:2003} color transformations.
All the isochrones agree very well with the data for $M_V\lesssim 8$, which correspond to $M \gtrsim 0.6\; M_\odot$.
At lower masses, however, the purely theoretical PHOENIX transformations produce the largest discrepancy in both CMDs. 
In the $(B-V)$ CMD, both semi-empirical transformations perform equally well in providing a good match of the data over the whole mass range considered here (i.e. $M\gtrsim 0.30 \, M_\odot$).
In the $(V-I)$ CMD, on the other hand, all the relations fail to reproduce satisfactorily the fainter magnitude end ($M_V \gtrsim 8.5$). 
The \citet{VandenBerg_Clem:2003} transformation results in a reasonable fit up to $M_V \lesssim 10$.

It is worth noting that \citet{Lejeune_ea:1998} caution that their empirical colors between $4500$ and $2000$ K have higher uncertainties than the ones at higher temperatures, due to the difficulties of defining a continuum for these stars. 
These results are also compatible with the discussion in section~3.4 of \citet{VandenBerg_Clem:2003}, who argued that the failure of the \citet{BCAH98} model to match the observations at the faint end of the $(V-I)$ CMD is not due to a flaw intrinsic to the models, but to the color-$T_{\rm eff}$ relations used (for the $M_V$ magnitude in particular), inaccurate for low mass stars. 
A similar result has been reported by \citet{Stauffer_ea:2007} about the models of both \citet{BCAH98} and \citet{Siess_ea:2000}.

The disagreement between models and observations for $T_{\rm eff} \lesssim 4000$ K is usually ascribed to the existence of sources of opacity unaccounted for in atmosphere and/or interior models.
Moreover, for a young cluster as the Pleiades, and up to about $500$ Myr, stars in this range of $T_{\rm eff}$ are still in the PMS/early ZAMS phase.
The missing opacities problem, combined with the notorious uncertainties of PMS models, can also lead to a disagreement between cluster ages estimated on the basis of MS and PMS isochrones, \citep{Bell_ea:2012}.
A detailed analysis of PMS evolution is deferred to future work.

\bibliographystyle{mn2e}

\begin{thebibliography}{}

\bibitem[Allard et al.(1997)]{Allard_ea:1997} Allard, F., Hauschildt, P.~H., Alexander, D.~R., \& Starrfield, S.\ 1997, \araa, 35, 137 

\bibitem[Allard et al.(2011)]{Allard_ea:2011} Allard, F., Homeier, D., \& Freytag, B.\ 2011, 16th Cambridge Workshop on Cool Stars, Stellar Systems, and the Sun, 448, 91 

\bibitem[Andersen et al.(1988)]{Andersen_ea:1988} Andersen, J., Clausen, J.~V., Nordstrom, B., Gustafsson, B., \& Vandenberg, D.~A.\ 1988, \aap, 196, 128 

\bibitem[Asplund et al.(2009)]{Asplund_ea:2009} Asplund, M., Grevesse, N., Sauval, A.~J., \& Scott, P.\ 2009, \araa, 47, 481 

\bibitem[Bahcall \& Pinsonneault(1992)]{Bahcall_Pinsonneault:1992} Bahcall, J.~N., \& Pinsonneault, M.~H.\ 1992, \apjl, 395, L119 

\bibitem[Baraffe et al.(1998)]{BCAH98} Baraffe, I., Chabrier, G., Allard, F., \& Hauschildt, P.~H.\ 1998, \aap, 337, 403 

\bibitem[Barnes(2010)]{Barnes:2010} Barnes, S.~A.\ 2010, \apj, 722, 222 

\bibitem[Barnes \& Kim(2010)]{Barnes_Kim:2010} Barnes, S.~A., \& Kim, Y.-C.\ 2010, \apj, 721, 675 

\bibitem[Basu \& Antia(2008)]{Basu_Antia:2008} Basu, S., \& Antia, H.~M.\ 2008, \physrep, 457, 217 

\bibitem[Bell et al.(2012)]{Bell_ea:2012} Bell, C.~P.~M., Naylor, T., Mayne, N.~J., Jeffries, R.~D., \& Littlefair, S.~P.\ 2012, \mnras, 424, 3178

\bibitem[Berger et al.(2006)]{Berger_ea:2006} Berger, D.~H., Gies, D.~R., McAlister, H.~A., et al.\ 2006, \apj, 644, 475 

\bibitem[Bertelli et al.(2008)]{Bertelli_ea:2008} Bertelli, G., Girardi, L., Marigo, P., \& Nasi, E.\ 2008, \aap, 484, 815 

\bibitem[B{\"o}hm-Vitense(1958)]{BV58} B{\"o}hm-Vitense, E.\ 1958, Zs. Ap., 46, 108 

\bibitem[Bonaca et al.(2012)]{Bonaca_ea:2012} Bonaca, A., Tanner, J.~D., Basu, S., et al.\ 2012, \apjl, 755, L12 

\bibitem[Boyajian et al.(2012a)]{Boyajian_ea:2012a} Boyajian, T.~S., McAlister, H.~A., van Belle, G., et al.\ 2012, \apj, 746, 101 

\bibitem[Boyajian et al.(2012b)]{Boyajian_ea:2012b} Boyajian, T.~S., von Braun, K., van Belle, G., et al.\ 2012, \apj, 757, 112

\bibitem[Bressan et al.(2012)]{Bressan_ea:2012} Bressan, A., Marigo, P., Girardi, L., et al.\ 2012, \mnras, 427, 127 

\bibitem[Carter et al.(2011)]{Carter_ea:2011} Carter, J.~A., Fabrycky, D.~C., Ragozzine, D., et al.\ 2011, Science, 331, 562 

\bibitem[Chabrier et al.(1996)]{Chabrier_ea:1996} Chabrier, G., Baraffe, I., \& Plez, B.\ 1996, \apjl, 459, L91

\bibitem[Chabrier \& Baraffe(1997)]{Chabrier_Baraffe:1997} Chabrier, G., \& Baraffe, I.\ 1997, \aap, 327, 1039

\bibitem[Chabrier et al.(2007)]{Chabrier_ea:2007} Chabrier, G., Gallardo, J., \& Baraffe, I.\ 2007, \aap, 472, L17 

\bibitem[Chan \& Sofia(1989)]{Chan_Sofia:1989} Chan, K. L., \& Sofia, S.\ 1989, \apj, 336, 1022 

\bibitem[Charbonneau (2013)]{Charbonneau:2013} Charbonneau, P. \ 2013, Society for Astronomical Sciences Annual Symposium, 39, 187

\bibitem[Christensen-Dalsgaard(1997)]{JCD97} Christensen-Dalsgaard, J.\ 1997, SCORe'96 : Solar Convection and Oscillations and their Relationship, 225, 3 

\bibitem[Cyburt et al.(2008)]{Cyburt_ea:2008} Cyburt, R.~H., Fields, B.~D., \& Olive, K.~A.\ 2008, JCAP, 11, 12 

\bibitem[Daeppen et al.(1988)]{Dappen_ea:1988} Daeppen, W., Mihalas, D., Hummer, D.~G., \& Mihalas, B.~W.\ 1988, \apj, 332, 261 

\bibitem[Demarque et al.(2004)]{Demarque_ea:2004} Demarque, P., Woo, J.-H., Kim, Y.-C., \& Yi, S.~K.\ 2004, \apjs, 155, 667 

\bibitem[Demarque et al.(2008)]{Demarque_ea:2008} Demarque, P., Guenther, D.~B., Li, L.~H., Mazumdar, A., \& Straka, C.~W.\ 2008, \apss, 316, 31 

\bibitem[Demory et al.(2009)]{Demory_ea:2009} Demory, B.-O., S{\'e}gransan, D., Forveille, T., et al.\ 2009, \aap, 505, 205 

\bibitem[Dotter et al.(2007)]{Dotter_ea:2007} Dotter, A., Chaboyer, B., Jevremovi{\'c}, D., et al.\ 2007, \aj, 134, 376 

\bibitem[Dotter et al.(2008)]{Dotter_ea:2008} Dotter, A., Chaboyer, B., Jevremovi{\'c}, D., et al.\ 2008, \apjs, 178, 89 

\bibitem[Doyle et al.(2011)]{Doyle_ea:2011} Doyle, L.~R., Carter, J.~A., Fabrycky, D.~C., et al.\ 2011, Science, 333, 1602 

\bibitem[Feiden \& Chaboyer(2012)]{Feiden_Chaboyer:2012} Feiden, G.~A., \& Chaboyer, B.\ 2012, \apj, 757, 42 

\bibitem[Feiden \& Chaboyer(2012b)]{Feiden_Chaboyer:2012b} Feiden, G.~A., \& Chaboyer, B.\ 2012, \apj, 761, 30 

\bibitem[Ferguson et al.(2005)]{Ferguson_ea:2005} Ferguson, J.~W., Alexander, D.~R., Allard, F., Barman, T., Bodnarik, J.~G., Hauschildt, P.~H., Heffner-Wong, A., \& Tamanai, A.\ 2005, \apj, 623, 585 


\bibitem[Gough \& Tayler(1966)]{Gough_Tayler:1966} Gough, D.~O., \& Tayler, R.~J.\ 1966, \mnras, 133, 85 

\bibitem[Grevesse \& Noels(1993)]{Grevesse_Noels:1993} Grevesse, N., \& Noels, A.\ 1993, Origin and Evolution of the Elements, 15 

\bibitem[Grevesse \& Sauval(1998)]{Grevesse_Sauval:1998} Grevesse, N., \& Sauval, A.~J.\ 1998, \ssr, 85, 161 

\bibitem[Hauschildt et al.(1999)]{Hauschildt_ea:1999} Hauschildt, P.~H., Allard, F., \& Baron, E.\ 1999, \apj, 512, 377 

\bibitem[Henry \& McCarthy(1993)]{Henry_McCarthy:1993} Henry, T.~J., \& McCarthy, D.~W., Jr.\ 1993, \aj, 106, 773 

\bibitem[Henry(2004)]{Henry:2004} Henry, T.~J.\ 2004, Spectroscopically and Spatially Resolving the Components of the Close Binary Stars, 318, 159 

\bibitem[Henry et al.(2006)]{Henry_ea:2006} Henry, T.~J., Jao, W.-C., Subasavage, J.~P., et al.\ 2006, \aj, 132, 2360 

\bibitem[Hoxie(1973)]{Hoxie:1973} Hoxie, D.~T.\ 1973, \aap, 26, 437 

\bibitem[Hummer \& Mihalas(1988)]{Hummer_Mihalas:1988} Hummer, D.~G., \& Mihalas, D.\ 1988, \apj, 331, 794 

\bibitem[Iglesias \& Rogers(1996)]{Iglesias_Rogers:1996} Iglesias, C.~A., \& Rogers, F.~J.\ 1996, \apj, 464, 943 

\bibitem[Kasting et al.(1993)]{Kasting_ea:1993} Kasting, J.~F., Whitmire, D.~P., \& Reynolds, R.~T.\ 1993, Icarus, 101, 108 

\bibitem[Kawaler(1988)]{Kawaler:1988} Kawaler, S.~D.\ 1988, \apj, 333, 236 

\bibitem[Kim et al. (1996)]{Kim_ea:1996} Kim, Y.-C., Fox, P.~A.,  Demarque, P. \& Sofia, S. \ 1996, \apj, 461, 499 

\bibitem[Kim \& Demarque(1996)]{Kim_Demarque:1996} Kim, Y.-C., \& Demarque, P.\ 1996, \apj, 457, 340 

\bibitem[Kim et al.(2002)]{Kim_ea:2002} Kim, Y.-C., Demarque, P., Yi, S.~K., \& Alexander, D.~R.\ 2002, \apjs, 143, 499 

\bibitem[Kopparapu et al.(2013)]{Kopparapu_ea:2013} Kopparapu, R.~K., Ramirez, R., Kasting, J.~F., et al.\ 2013, \apj, 765, 131 

\bibitem[Kraus et al.(2011)]{Kraus_ea:2011} Kraus, A.~L., Tucker, R.~A., Thompson, M.~I., Craine, E.~R., \& Hillenbrand, L.~A.\ 2011, \apj, 728, 48 

\bibitem[Krishna Swamy(1966)]{KrishnaSwamy:1966} Krishna Swamy, K.~S.\ 1966, \apj, 145, 174 

\bibitem[Lacy(1977)]{Lacy:1977} Lacy, C.~H.\ 1977, \apj, 218, 444 

\bibitem[Landin et al.(2010)]{Landin_ea:2010} Landin, N.~R., Mendes, L.~T.~S., \& Vaz, L.~P.~R.\ 2010, \aap,510, A46 

\bibitem[Lejeune et al.(1997)]{Lejeune_ea:1997} Lejeune, T., Cuisinier, F., \& Buser, R.\ 1997, \aaps, 125, 229 

\bibitem[Lejeune et al.(1998)]{Lejeune_ea:1998} Lejeune, T., Cuisinier, F., \& Buser, R.\ 1998, \aaps, 130, 65 

\bibitem[L{\'o}pez-Morales(2007)]{Lopez-Morales:2007} L{\'o}pez-Morales, M.\ 2007, \apj, 660, 732 

\bibitem[Ludwig et al. (1999)]{Ludwig_ea:1999} Ludwig, H.-G., Freytag, B. \& Steffen, M.\ 1999, \aap, 346, 111 

\bibitem[Lydon \& Sofia(1995)]{Lydon_Sofia:1995} Lydon, T.~J., \& Sofia, S.\ 1995, \apjs, 101, 357 

\bibitem[MacDonald \& Mullan(2012)]{MacDonald_Mullan:2012} MacDonald, J., \& Mullan, D.~J.\ 2012, \mnras, 421, 3084 

\bibitem[MacDonald \& Mullan(2013)]{MacDonald_Mullan:2013} MacDonald, J., \& Mullan, D.~J.\ 2013, \apj, 765, 126 

\bibitem[Meibom \& Mathieu(2005)]{Meibom_Mathieu:2005} Meibom, S., \& Mathieu, R.~D.\ 2005, \apj, 620, 970 

\bibitem[Mihalas et al.(1988)]{Mihalas_ea:1988} Mihalas, D., Dappen, W., \& Hummer, D.~G.\ 1988, \apj, 331, 815 

\bibitem[Morales et al.(2009)]{Morales_ea:2009} Morales, J.~C., Ribas, I., Jordi, C., et al.\ 2009, \apj, 691, 1400 

\bibitem[Morales et al.(2010)]{Morales_ea:2010} Morales, J.~C., Gallardo, J., Ribas, I., et al.\ 2010, \apj, 718, 502 

\bibitem[Noyes et al.(1984)]{Noyes_ea:1984} Noyes, R.~W., Weiss, N.~O., \& Vaughan, A.~H.\ 1984, \apj, 287, 769 

\bibitem[Reiners(2012)]{Reiners:2012} Reiners, A.\ 2012, Living Reviews in Solar Physics, 9, 1 

\bibitem[Ribas et al.(2008)]{Ribas_ea:2008} Ribas, I., Morales, J.~C., Jordi, C., et al.\ 2008, \memsai, 79, 562 

\bibitem[Rogers \& Nayfonov(2002)]{Rogers_Nayfonov:2002} Rogers, F.~J., \& Nayfonov, A.\ 2002, \apj, 576, 1064 

\bibitem[Saumon et al.(1995)]{Saumon_ea:1995} Saumon, D., Chabrier, G., \& van Horn, H.~M.\ 1995, \apjs, 99, 713 

\bibitem[Schatzman(1962)]{Schatzman:1962} Schatzman, E.\ 1962, Annales d'Astrophysique, 25, 18 

\bibitem[Schrijver \& Zwaan(2008)]{Schrijver_Zwaan:2008} Schrijver, C.~J., \& Zwaan, C.\ 2008, Solar and Stellar Magnetic Activity, Cambridge, UK: Cambridge University Press, 2008  

\bibitem[Siess et al.(2000)]{Siess_ea:2000} Siess, L., Dufour, E., \& Forestini, M.\ 2000, \aap, 358, 593 

\bibitem[Sills et al.(2000)]{Sills_ea:2000} Sills, A., Pinsonneault, M.~H., \& Terndrup, D.~M.\ 2000, \apj, 534, 335 

\bibitem[Skumanich(1972)]{Skumanich:1972} Skumanich, A.\ 1972, \apj, 171, 565 

\bibitem[Spada et al.(2010)]{Spada_ea:2010} Spada, F., Lanzafame, A.~C., \& Lanza, A.~F.\ 2010, \mnras, 404, 641 

\bibitem[Spada et al.(2011)]{Spada_ea:2011} Spada, F., Lanzafame, A.~C., Lanza, A.~F., Messina, S., \& Collier Cameron, A.\ 2011, \mnras, 416, 447 

\bibitem[Spada \& Demarque(2012)]{Spada_Demarque:2012} Spada, F., \& Demarque, P.\ 2012, \mnras, 422, 2255 

\bibitem[Stahler(1983)]{Stahler:1983} Stahler, S.~W.\ 1983, \apj, 274, 822 

\bibitem[Stauffer et al.(2003)]{Stauffer_ea:2003} Stauffer, J.~R., Jones, B.~F., Backman, D., et al.\ 2003, \aj, 126, 833 

\bibitem[Stauffer et al.(2007)]{Stauffer_ea:2007} Stauffer, J.~R., Hartmann, L.~W., Fazio, G.~G., et al.\ 2007, \apjs, 172, 663 

\bibitem[Tanner et al.(2013)]{Tanner_ea:2013} Tanner, J.~D., Basu, S., \& Demarque, P.\ 2013, \apj, 767, 78 

\bibitem[Tayler(1986)]{Tayler:1986} Tayler, R.~J.\ 1986, \mnras, 220, 793 

\bibitem[Thoul et al.(1994)]{Thoul_ea:1994} Thoul, A.~A., Bahcall, J.~N., \& Loeb, A.\ 1994, \apj, 421, 828 

\bibitem[Torres \& Ribas(2002)]{Torres_Ribas:2002} Torres, G., \& Ribas, I.\ 2002, \apj, 567, 1140 

\bibitem[Torres et al.(2010)]{Torres_ea:2010} Torres, G., Andersen, J., \& Gim{\'e}nez, A.\ 2010, \aapr, 18, 67 

\bibitem[Torres(2013)]{Torres:2013} Torres, G.\ 2013, Astronomische Nachrichten, 334, 4 

\bibitem[Trampedach et al.(2006)]{Trampedach_ea:2006} Trampedach, R., D{\"a}ppen, W., \& Baturin, V.~A.\ 2006, \apj, 646, 560 

\bibitem[VandenBerg \& Clem(2003)]{VandenBerg_Clem:2003} VandenBerg, D.~A., \& Clem, J.~L.\ 2003, \aj, 126, 778 

\bibitem[VandenBerg et al.(2006)]{VandenBerg_ea:2006} VandenBerg, D.~A., Bergbusch, P.~A., \& Dowler, P.~D.\ 2006, \apjs, 162, 375 

\bibitem[Winn et al.(2011)]{Winn_ea:2011} Winn, J.~N., Albrecht, S., Johnson, J.~A., et al.\ 2011, \apjl, 741, L1 

\bibitem[Wright et al.(2011)]{Wright_ea:2011} Wright, N.~J., Drake, J.~J., Mamajek, E.~E., \& Henry, G.~W.\ 2011, \apj, 743, 48 

\bibitem[Yi et al.(2001)]{Yi_ea:2001} Yi, S., Demarque, P., Kim, Y.-C., et al.\ 2001, \apjs, 136, 417 

\end{thebibliography}

\end{document}